\documentclass[%
 reprint,
superscriptaddress,
showpacs,preprintnumbers,
 amsmath,amssymb,
 aps,
]{revtex4-1}

\usepackage{graphicx}
\usepackage{dcolumn}
\usepackage{bm}

\newcommand\redout{\bgroup\markoverwith{\textcolor{red}{\rule[.5ex]{2pt}{0.4pt}}}\ULon}

\usepackage[colorlinks=true,citecolor=blue]{hyperref}
\hypersetup{colorlinks=true,citecolor=blue,linkcolor=red,urlcolor=blue}

\begin{document}
\title{Plasmons at the LaAlO$_3$/SrTiO$_3$ interface and in the graphene-LaAlO$_3$/SrTiO$_3$ double layer}
\author{A. Faridi}
\affiliation{Department of Physics, Sharif University of Technology, Tehran, Iran}
\affiliation{School of Physics, Institute for Research in Fundamental Sciences (IPM), Tehran 19395-5531, Iran}

\author{Reza Asgari}
\email{asgari@ipm.ir}
\affiliation{School of Physics, Institute for Research in Fundamental Sciences (IPM), Tehran 19395-5531, Iran}
\affiliation{School of Nano Science, Institute for Research in Fundamental Sciences (IPM), Tehran 19395-5531, Iran}

\date{\today}

\begin{abstract}
We study plasmon modes of the two-dimensional electron gas residing at the interface of band insulators $\rm{LaAlO_3}$ and $\rm{SrTiO_3}$  (LAO/STO) and the plasmon excitations of graphene-LAO/STO double layer as well. Considering the electron-electron interaction within random phase approximation, we calculate the plasmon dispersions of both systems numerically and in the long wavelength limit analytical expressions for collective modes are found. One optical mode and two (three) acoustic modes are predicted for the LAO/STO (graphene-LAO/STO) system where only the uppermost acoustic mode of both systems can emerge above the electron-hole continuum depending on the characteristics of each system. In the case of LAO/STO interface, thanks to the spatial separation between $\rm{t_{2g}}$ orbitals, the upper acoustic mode might be undamped at the long wavelength limit depending on the exact value of the dielectric constant of $\rm{SrTiO_3}$. Same as other double layer systems, the interlayer distance for the graphene-LAO/STO system plays a crucial role in damping the upper acoustic mode. Faster damping of all plasmon modes of the present double layer system in comparison with the ones with conventional two-dimensional electron gas instead of $\rm{t_{2g}}$ electron gas is also found
due to heavier effective masses of the gas and also stronger interlayer Coulomb interaction.
\end{abstract}


\maketitle

\section{Introduction}

Plasmon modes represent a second kind of possible elementary excitation
for the Fermi liquid~\cite{pines, vignale}. Basically, they involve a cooperative motion of the
system, governed by the global interaction between the electrons.

The collective modes differ in nature from the individual
excitations where an individual component acts to force the motion of all the other particles.
The innermost of the excitation is a single quasiparticle where it is surrounded
by an appropriate polarization cloud. In a collective
mode, by contrast, all particles play an equal role in such a way that the distribution function
extends smoothly over the entire Fermi surface. Plasmon modes can be observed by a variety of experimental tools including inelastic light scattering~\cite{pellegrini_review_2006}, which has been widely used to probe plasmons in semiconductor heterostructures~\cite{hirjibehedin_prb_2007,kainth_prb_1999,tovstonog_prb_2002},
but also by surface-physics techniques like high-resolution electron-energy-loss spectroscopy~\cite{liu_prb_2008}, more indirectly, angle-resolved photoemission spectroscopy~\cite{eeinteractionsgraphene}, double-layer field-effect transistors with a grating gate~\cite{peralta_apl_2002} and scattering-type scanning near-field optical microscopy~\cite{graphene} can also be used to detect plasmons.

The physics of decoupled two-dimensional (2D) electron systems
has been a subject of theoretical and experimental interest since it was
recognized~\cite{pogrebinskii_1977,price_physicaB_1983} that electron-electron interactions
allow the energy and momentum to
be transferred between layers while maintaining separate particle-number conservation.
The study of Coulomb-coupled 2D systems has now been revitalized
by advances which have made it possible to prepare robust and ambipolar 2DEGs,
based on graphene~\cite{graphenereviews} which are described by an ultrarelativistic wave equation instead of the non-relativistic Schr\"{o}dinger equation.

The discovery of two-dimensional electron gas (2DEG) at the interface
between two perovskite band insulators $\rm{LaAlO_3}$ (LAO) and
$\rm{SrTiO_3}$ (STO) when more than $3$ unit cells of LAO
are grown on a TiO2-terminated STO crystal has attracted attention. Electrons at the interfaces with partially occupied d orbitals interact with each other and with the lattice. This gives rise to wide electronic properties, such as conductivity~\cite{ohtomo2,thiel}
superconductivity~\cite{Reyren} and magnetism~\cite{Brinkman}. This
2DEG is located only a few nm below the surface and
its properties are therefore sensitive to other
materials deposited on this surface. More interestingly, decoupled structures
combining graphene with LAO/STO junctions represent
an exciting platform in which novel phenomena may
emerge from the strong electronic coupling of the respective
2DEGs. Recently, the transport properties of hybrid devices obtained by depositing graphene on
a LAO/STO oxide junction hosting a 4 nm-deep 2DEG has been studied~\cite{Huang, Marco}.

The existence of plasmon modes at the interface of  $\rm LaAlO_3$ and  $\rm SrTiO_3$ has been confirmed experimentally~\cite{ruotsa} and was predicted theoretically~\cite{Park}. In this work, we study collective modes of the 2DEG residing at this interface and the plasmon excitations of graphene-LAO/STO double layer as well by using a simple three-band model Hamiltonian for 2DEG at the oxide interface. Having considered electron-electron interaction within random phase approximation, we calculate the plasmon dispersions of both systems numerically. One optical mode and two (three) acoustic modes are predicted for the LAO/STO (graphene-LAO/STO) system where only the uppermost acoustic mode of both systems can emerge above the electron-hole continuum depending on the characteristics of each system. We also derive analytical expressions for both optical and acoustic (damped and undamped) collective modes at the long wavelength limit which were in perfect agreement with the numerical findings. In the case of LAO/STO system, while the low-lying acoustic mode is always damped, the emergence of the upper acoustic mode above the particle-hole excitations of the system, depends on the dielectric constant of surrounding medium as well as the carrier density at the interface. More precisely, for a large dielectric constant of the medium, not only the uppermost acoustic mode of the system disappears, even at the long wavelength limit, but also the damping of the optical mode begins in a lower energy and longer wavelength. Because the dielectric constant of $\rm SrTiO_3$ increases with decreasing the carrier density at the interface~\cite{copie,gariglio} and leads to a strong suppression of the electron-electron interaction, it is believed that plasmon modes can not emerge in such a system. This could be the main reason why in a recent work, where a polaronic metal state has been found, the plasmon modes were not detected~\cite{cancellieri}. It is also interesting to know that the polaronic state breaks down for large electron densities~\cite{wang}. The detection of the plasmon modes, on the other hand, depends strongly on the cleanliness of the system, because defects and impurities can accelerate plasmon damping. Apparently, the experimental system in \cite{cancellieri} can not be considered as a clean system since the electron mobility reported in this paper is of order of $1000~ \rm{cm^2(Vs)^{-1}}$ at low temperature which is almost $10$ times smaller than the typical mobility of this system~\cite{ohtomo2,kalabukhov,herranz,sulpizio}. Therefore, a high-quality sample with high electron concentrations is needed to observe plasmon modes for such a system.

In the case of graphene-LAO/STO,  we also find an expression for the critical value of the interlayer distance above which the upper acoustic mode can emerge in the region where the imaginary part of the dielectric function of the system vanishes and the mode is undamped at least in the long wavelength limit.

This manuscript is organized as follows. In Sect.~II we present the model and results we have used to introduce the linear-response functions which describe collective electron dynamics and to describe plasmon modes of 2DEG at the transition metal oxide interface.
In Sect.~III we present our main analytical and numerical results for the dispersion of optical and acoustic plasmons in the decoupled graphene-LAO/STO system. Finally, Sect.~IV contains a summary of our main conclusions.

\section{Plasmon modes of LAO/STO}

It has been illustrated that $\rm t_{2g}$ orbitals ($\rm {d_{xy},d_{xz},d_{yz}}$) of Ti atom host the 2DEG at the interface of $\rm SrTiO_3$ and $\rm LaAlO_3$ \cite{pentcheva, zoran}. The larger spatial extension of the $\rm d_{xy}$ ($\rm{d_{yz}/d_{xz}}$) orbital in $\hat{x}$ ($\hat{y}$/$\hat{x}$) and $\hat{y}$ ($\hat{z}$/$\hat{z}$) directions with respect to $\hat{z}$ ($\hat{x}$/$\hat{y}$) results in the formation of an isotropic $\rm d_{xy}$ band with the same large hopping, $t$ and  light effective mass, $m_L$ in both $\hat{x}$ and $\hat{y}$ directions and two anisotropic $\rm d_{yz}$ and $\rm d_{xz}$ bands with a large hopping, $t$ and light effective mass, $m_L$ in $\hat{y} (\hat{x})$ direction and small hopping, $t'$ and heavy effective mass, $m_H$ in $\hat{x} (\hat{y})$ direction at the interface. Furthermore, the confinement of 2DEG along $\hat{z}$ lowers the energy of $\rm d_{xy}$ band by the off-set energy $\Delta$ with respect to the other two bands.  These considerations lead to a tight-binding Hamiltonian, diagonal in orbital space with elements $[\hat{\cal {H}}_0]_{ij}=(t_{x,i}\cos kx+t_{y,i}\cos ky)\delta_{ij}$ where $t_{x,2}=t_{y,3}=t'$ and $t_{x,1}=t_{y,1}=t_{y,2}=t_{x,3}=t$~\cite{khalsa,khalsa2,zhong}. Expanding the tight-binding Hamiltonian around the $\Gamma$ point, a minimal model Hamiltonian of the carriers at the interface of $\rm LaAlO_3$ and $\rm SrTiO_3$ can be written as~\cite{nayak,tolsma,tolsma2,Faridi}
\begin{equation}
 \begin{split}\label{eq:fullH1}
{\hat {\cal H}} &= \sum_{ k,\theta,i}( \frac{\hbar^2 { k}^2\cos^2(\theta)}{2m_{x,i}}+\frac{\hbar^2 { k}^2\sin^2(\theta)}{2m_{y,i}}+\Delta_{i}){\hat \phi}^\dagger_{\bm k, i}{\hat \phi}_{\bm k, i}\\
&+\frac{1}{2 S}\sum_{{\bm q} \neq {\bm 0},i, j} V_{ij}(q) {\hat \rho}^{({i})}_{\bm q} {\hat \rho}^{({j})}_{-{\bm q}}
\end{split}
\end{equation}

The first term of the Hamiltonian is the kinetic part where $i=1,2$ and $3$ corresponds to $\rm d_{xy},d_{yz}$ and $d_{xz}$ bands, $\Delta_1=0$, $\Delta_2=\Delta_3=\Delta$ and $\hat{\phi}^\dagger_{\bm{k}, i}$($\hat{\phi}_{\bm{k},i}$) creation (annihilation) operator of an electron with momentum $\bm{k}$ in band $i$. We also emphasize that $m_{x,1}=m_{y,1}=m_{y,2}=m_{x,3}=m_L$ and $m_{x,2}=m_{y,3}=m_H$. The second term indicates the long-range Coulomb interaction in the system, where $V_{ij}(q)=v(q)e^{-qa_{ij}}=\frac{2\pi e^2}{\bar{\epsilon} q}e^{-qa_{ij}}$ is the 2D Coulomb electron-electron interaction and  $\bar{\epsilon}$ is the average dielectric constant of the surrounding medium, $S$ is the sample area and  ${\hat \rho}^{({i})}_{\bm q} = \sum_{{\bm k}} {\hat \phi}^\dagger_{{\bm k} - {\bm q},  i}{\hat \phi}_{{\bm k},  i}$ is the density operator of band $i$. Because of the larger effective mass of $\rm d_{xy}$ orbital in $\hat{z}$ direction, it is more confined to the interface which results in an orbital dependence of the Coulomb interaction~\cite{khalsa, tolsma}. This effect which plays a crucial role in determining the second collective mode of the system is captured by introducing an effective distance $a_{ij}$ between orbital $i$ and orbital $j$ and vanishes except for $a_{1,2}=a_{2,1}=a_{1,3}=a_{3,1}=a$.

It is important to mention that the spin-orbit interaction also plays a role in this system. Furthermore, the inversion symmetry breaking along $\hat{z}$ at the interface leads to the Rashba interaction as well~\cite{nayak,khalsa2,zhong,zhou1}. The most significant effects of these terms on the band dispersion emerge near the band degeneracy points (in the form of orbital mixing and also band splitting). Since in this paper, we work with a density regime high enough to stay far from these points, accordingly, the tight-binding Hamiltonian of Eq.~\eqref{eq:fullH1} can properly describe the system~\cite{tolsma,tolsma2,Faridi}.

In order to find the plasmon modes, we need to find the poles of the linear-response function of the system which within the random phase approximation (RPA) is given by~\cite{vignale}
\begin{equation}\label{eq:xi}
\boldsymbol{\hat{\chi}}^{-1}(\mathbf{q,\omega})=\boldsymbol{\hat{\chi^0}^{-1}}(\mathbf{q,\omega})-\hat{V}_\mathbf{q}
\end{equation}
where $\boldsymbol{\hat{\chi}^0}(\mathbf{q,\omega})$ is the noninteracting density-density response matrix of the system which is diagonal with elements ${\chi^0_1}({q,\omega})$, the well-known noninteracting density-density response function of the conventional 2DEG~\cite{vignale}, ${\chi^0_2}(\mathbf{q,\omega})$ and ${\chi^0_3}(\mathbf{q,\omega})$ the noninteracting density-density response functions of a two-dimensional electron gas with elliptical band dispersion. While the first element ${\chi^0_1}(\mathbf{q,\omega})$ is already known, the response function of elliptical band dispersion can be easily found by applying a Herring-Vogt transformation~\cite{herring} $ k_x\to k_x\sqrt{m_x/m_D}$ and $k_y\to k_y\sqrt{m_y/m_D}$ with $m_D=\sqrt{m_xm_y}$ and so we will have~\cite{tolsma, Faridi}
\begin{equation}\label{eq:chi02}
 \chi^0_{2}(\mathbf{q},\omega)=\chi^0_1(q',\omega;m_D)\mid_{q'\to \big(q_x^2\sqrt{\xi}+q_y^2\sqrt{\frac{1}{\xi}}\big)^{1/2}}
\end{equation}
and
\begin{equation}\label{eq:chi03}
 \chi^0_{3}(\mathbf{q},\omega)=\chi^0_1(q',\omega;m_D)\mid_{q'\to \big(q_x^2\sqrt{\frac{1}{\xi}}+q_y^2\sqrt{\xi}\big)^{1/2}}
\end{equation}
where $\xi=m_L/m_H$. Also making use of this transformation one can define an identical parameter $k_{\rm F,2}=\sqrt{2m_D\epsilon_{\rm F}}/\hbar$ related to the Fermi wave vector average for both elliptical bands. Note that the density-density response function of the circular band, $\chi^0_{1}(q,\omega)$ depends only on $|\mathbf{q}|\equiv q$.

The collective modes of the system can be achieved by the poles of the response function Eq.~\eqref{eq:xi} or equivalently the zeros of the dielectric function
\begin{equation}\label{eq:eps}
 \begin{split}
 \varepsilon(\mathbf{q},\omega)&=(1-v(q)\chi_1^0({q},\omega))[1-v(q)(\chi_2^0(\mathbf{q},\omega)+\chi_3^0(\mathbf{q},\omega))]\\
&-v^2(q)e^{-2qa}\chi_1^0({q},\omega)[\chi_2^0(\mathbf{q},\omega)+\chi_3^0(\mathbf{q},\omega)]
 \end{split}
 \end{equation}

Since in this system the Fermi surface is crossed by three bands with different Fermi velocities, we expect to have three collective modes~\cite{dassarma2}; one optical plasmon mode with the square root dispersion relation which can be interpreted as the in-phase oscillations of the electrons of all bands and two acoustic plasmon modes which are the oscillations of the slower electrons of the system screened and damped by single-particle excitations of the faster carriers.
The optical plasmon mode always occurs above the electron-hole continuum of the fastest carriers of the system where the imaginary part of the dielectric function is zero. We note that in a 2DEG system the upper boundary of the electron-hole excitation region is defined as $\omega_+=v_{\rm F}|\mathbf{q}|$ in the long wavelength limit~\cite{vignale} where $v_{\rm F}$ is the electron Fermi velocity. The same expression can be used to describe the upper boundary of the electron-hole continuum of the elliptical bands using the transformations of Eqs.~\eqref{eq:chi02} and \eqref{eq:chi03} for $|\mathbf{q}|$. In this way we can define a  direction dependent upper electron-hole continuum boundary for the elliptical bands of the form
\begin{equation}\label{eq:v2}
\omega_{+, yz}(\theta)=(\hbar k_{{\rm F} ,2}/m_D)(\cos^2(\theta)\sqrt{\xi}+\sin^2(\theta)\sqrt{1/\xi})^{1/2}|\mathbf{q}|
\end{equation}
and
\begin{equation}\label{eq:v3}
\omega_{+, xz}(\theta)=(\hbar k_{{\rm F} ,2}/m_D)(\cos^2(\theta)\sqrt{1/\xi}+\sin^2(\theta)\sqrt{\xi})^{1/2}|\mathbf{q}|
\end{equation}

{\color{red}}Showing the direction independent part of the above boundaries with $v_{{\rm F}2}=\hbar k_{{\rm F} ,2}/m_D$ and the angle dependent parts as $z_1(\theta)$ and $z_2(\theta)$ , we can briefly write $\omega_{+ ,yz}(\theta)=v_{{\rm F}2}z_1(\theta)|\mathbf{q}|$ and $\omega_{+ ,xz}(\theta)=v_{{\rm F} 2}z_2(\theta)|\mathbf{q}|$. Note that the Fermi velocity of the circular band is simply $v_{{\rm F} ,xy}=v_{{\rm F}1}=\hbar k_{{\rm F}1}/m_L$ and it's electron-hole continuum boundary is $\omega_{+, xy}=v_{{\rm F} 1}|\mathbf{q}|$. For all electron densities used in this article and all orientations of $\mathbf{q}$, the velocity of the electrons of the circular band is larger than the other two bands.

\subsection{Analytical results at long wavelength limit}
At the long wavelength limit $q\to 0$ we would have an undamped optical collective mode of the form $\omega_{op}(q\to0)\propto\sqrt{q}$. To derive an analytic expression for the long wavelength limit of the optical plasmon dispersion of the system, we make use of the expansion of the noninteracting density-density response function of the 2DEG for $q\to 0$ and $\omega\gg qv_{F1}$ ($v_{F1}$ is the largest Fermi velocity among the carriers of all bands which as mentioned before, belongs to circular band here), which to the leading order in q is~\cite{vignale}
\begin{equation}\label{eq:chi0app}
 \chi^0_{i}(\mathbf{q},\omega)\simeq \frac{n_i|\mathbf{q}|^2}{m_i\omega^2}
\end{equation}
where $n_i$ and $m_i$ are the density and effective mass of the band $i$. We note that for elliptical bands $m_{2,3}=m_D$ we should apply the rescaling of Eqs.~\eqref{eq:chi02} \eqref{eq:chi03} as well.
Substituting Eq.~\eqref{eq:chi0app} for each band in Eq.~\eqref{eq:eps}, the frequency of the optical mode of the system in the long wavelength limit will be
\begin{equation}\label{eq:w_op1}
\omega_{op}^2(q\to0)=\frac{2\pi e^2}{\bar{\epsilon} }\Bigl[\frac{n_1}{m_L}+n_2( \frac{1}{m_L}+\frac{1}{m_H})\Bigr] q
\end{equation}

It can be seen that the optical mode of the system is the combination of optical mode of each band if treated separately.

The long wavelength behavior of the other two acoustic modes is of the type $\omega_{ac}(q\to0)=c_s q$ where $c_s$ is the acoustic mode group velocity. The first acoustic mode of the system occurs naturally in the region $qv_{F2}z_{m}(\theta)<\omega<qv_{F1}$($z_{m}=max\lbrace z_1,z_2\rbrace$) where the acoustic plasmon of the second and third bands are completely Landau damped by the electrons of the first band (by the first band we mean the band with the fastest electrons and the second and third bands are the bands with intermediate and low velocities)~\cite{takada}.
In order to calculate the damped plasmon modes we let the frequencies to have an imaginary part and solve $\varepsilon (\mathbf{q},\omega+i\delta)=0$. Following Ref.~\cite{santoro}, we are going to find an analytical expression for the oscillation velocity of the first acoustic mode of the system in the long wavelength limit $(q\to0)$. We introduce the power expansion
\begin{equation}
\omega_{ac}=(c_s+i\delta)q+c_2q^2+c_3q^3+\dotsm
\end{equation}
where coefficients $c_i $ can also be complex.

Substituting this expansion in Eq.~\eqref{eq:eps} and also in analytical expressions of the density-density response functions of the bands, we arrive at a Laurent-Taylor expansion for Eq.\eqref{eq:eps}
\begin{equation}
\varepsilon(\mathbf{q},(c_s+i\delta)q+c_2q^2+\dotsm)=f_{-1}q^{-1}+f_0+f_1q+\dotsm
\end{equation}
To find the zeros of the above equation, all the coefficients $f_i$ should vanish independently, which in our case are complex expressions . The only complex equation we need to solve for finding $c_s$ and $\delta$ is $f_{-1}=0$. (because $f_{-1}$ depends only on  $c_s$ and $\delta$). Therefore, we get two non-linear equations to solve in $\delta$ and $c_s$.
We arrive after some algebra to the following expression for $\delta$
\begin{equation}\label{eq:delta}
\delta=-\frac{(BC)\Bigl[\sqrt{BC}(1+4k_2a)-2k_2ac_s(\sqrt{B}+\sqrt{C})\Bigr]}{(\sqrt{A}v_{F2}c_s)(m_D/m_L+2k_2a)(z_1C^{3/2}+z_2B^{3/2})}
\end{equation}
where $k_2=2m_De^2/\bar{\epsilon} \hbar^2 $ is the Thomas-Fermi wave vector of the second and third bands. We also define $A=(v_{{\rm F}1}^2-c_s^2)$,  $B=(c_s^2-z_1^2v_{{\rm F}2}^2)$ and $C=(c_s^2-z_2^2v_{{\rm F}2}^2)$.

Substituting expression \eqref{eq:delta} in
\begin{equation}\label{eq:cs1}
 \begin{split}
&(2k_2a\delta)\Bigl[BCv_{F1}(\sqrt{B}+\sqrt{C})-Av_{F2}(z_1C^{3/2}+z_2B^{3/2})\Bigr]c_s^3\\
&-(BC)^{3/2}(v_{F1}\delta)(1+4k_2a) c_s^2\\
&-(A^{3/2}BC)(\frac{m_D}{m_L}+2k_2a)(\sqrt{B}+\sqrt{C})c_s\\
&+(ABC)^{3/2}(1+2\frac{m_D}{m_L}+4k_2a)=0
 \end{split}
 \end{equation}
we end up with an equation only in $c_s$ to be solved. Finding $c_s$ via this equation and putting it back in expression \eqref{eq:delta}, we will find $\delta$ as well. Notice that $c_s$ and $\delta$ are $\theta$-dependent and thus we expect that the plasmon modes change along different ${\bf q}$ directions. In order to have a physical plasmon mode, $\delta$ has to be very small.

It is also worthwhile to note that depending on the characteristics of the system, the first acoustic plasmon can also occur in the region $\omega>qv_{F1}$, so that it will be undamped unless it touches the boundary of the particle-hole continuum of the fastest electrons ($\omega=qv_{F1}$). In this case the acoustic plasmon group velocity can be obtained by the same procedure described before, except that the solutions of Eq.~\eqref{eq:eps} are real, $\delta=0$. Again, we find the following equation whose solution gives us the group velocity of the undamped acoustic plasmon
\begin{equation}\label{eq:cs2}
 \begin{split}
&(2k_2a)(\sqrt{B}+\sqrt{C}) c_s^2-\sqrt{A'}(\sqrt{B}+\sqrt{C})(\frac{m_D}{m_L}+2k_2a)c_s \\
&-\sqrt{BC}(1+4k_2a) c_s+\sqrt{A'BC}(1+2\frac{m_D}{m_L}+4k_2a)=0\\
 \end{split}
\end{equation}
With $A'=(c_s^2-v_{F1}^2)$. The threshold of $c_s$ above which the undamped acoustic plasmon emerges is $c_s=v_{F1}$. Using this value in either Eq.~\eqref{eq:cs1} or Eq.~\eqref{eq:cs2}, we find a critical value for $(ak_2)_{cr}$
\begin{equation}\label{eq:critical}
(ak_2)_{cr}=\frac{1}{2}\frac{\sqrt{BC}}{(\sqrt{B}+\sqrt{C})v_{F1}-2\sqrt{BC}}
\end{equation}

For larger values of $ak_2$ the acoustic plasmon occurs above the particle-hole continuum of all bands and for smaller values, the acoustic plasmon acquires a finite lifetime even at a large wavelength limit. For bilayer systems in which the distance between layers is arbitrary and all the characteristics of the system such as dielectric constant and effective masses are well-known, $(ak_2)_{cr}$ reduces to $a_{cr}$ which means that we can adopt the distance between layers so that the acoustic plasmon becomes an undamped mode. However, in the system in question, the effective distance between $\rm{d_{xy}}$ orbital and the other two orbitals are set by the density of carriers and as we increase the electron density at the interface, this effective distance reduces. Furthermore, the dielectric constant of $\rm{SrTiO_3}$ is not yet well-known. Although it has a large value of about $\epsilon_s=25000$ at low temperatures, it strongly depends on the electric field and the density of electrons at the interface. While scientists have proposed some expressions which can express the overall behavior of $\epsilon_s$~\cite{copie,gariglio}, it's exact value, specially at low temperatures, is still under debate. Therefore, to cover all possible situations in which collective modes can occur, we solve the problem for different values of $\epsilon_s$.

As we stated before, we have two acoustic plasmon modes in this system. The second acoustic mode lies in the region $qv_{F2}z_s(\theta)<\omega<qv_{F2}z_m(\theta)$ with $z_s=min\lbrace z_1,z_2 \rbrace$. In a similar way we obtained Eq.~\eqref{eq:delta} and Eq.~\eqref{eq:cs1}, we have found analytical expressions for the plasmon group velocity of the second acoustic mode and it's damping in Appendix~\ref{Sect:AppendixA}. This plasmon mode is always damped, even at the long wavelength limit. To find critical characteristics of the system above which the plasmon mode enters the second region we get the condition $(ak_2)_{cr}=-\frac{1}{2}\frac{m_D}{m_L}$ which is obviously impossible, therefore the second plasmon always occurs in the third region and it is strongly damped. This mode also disappears for $\theta=\pi/4$. In this case $v_{F,xz}=v_{F,yz}$ and the electron-hole continua of $\rm{d_{xz}}$ and $\rm{d_{yz}}$ bands coincide which means that the carriers of these bands behave completely the same and we have effectively a two-carrier system.

\begin{figure}[h]
\centering
      \includegraphics[width=1.\linewidth] {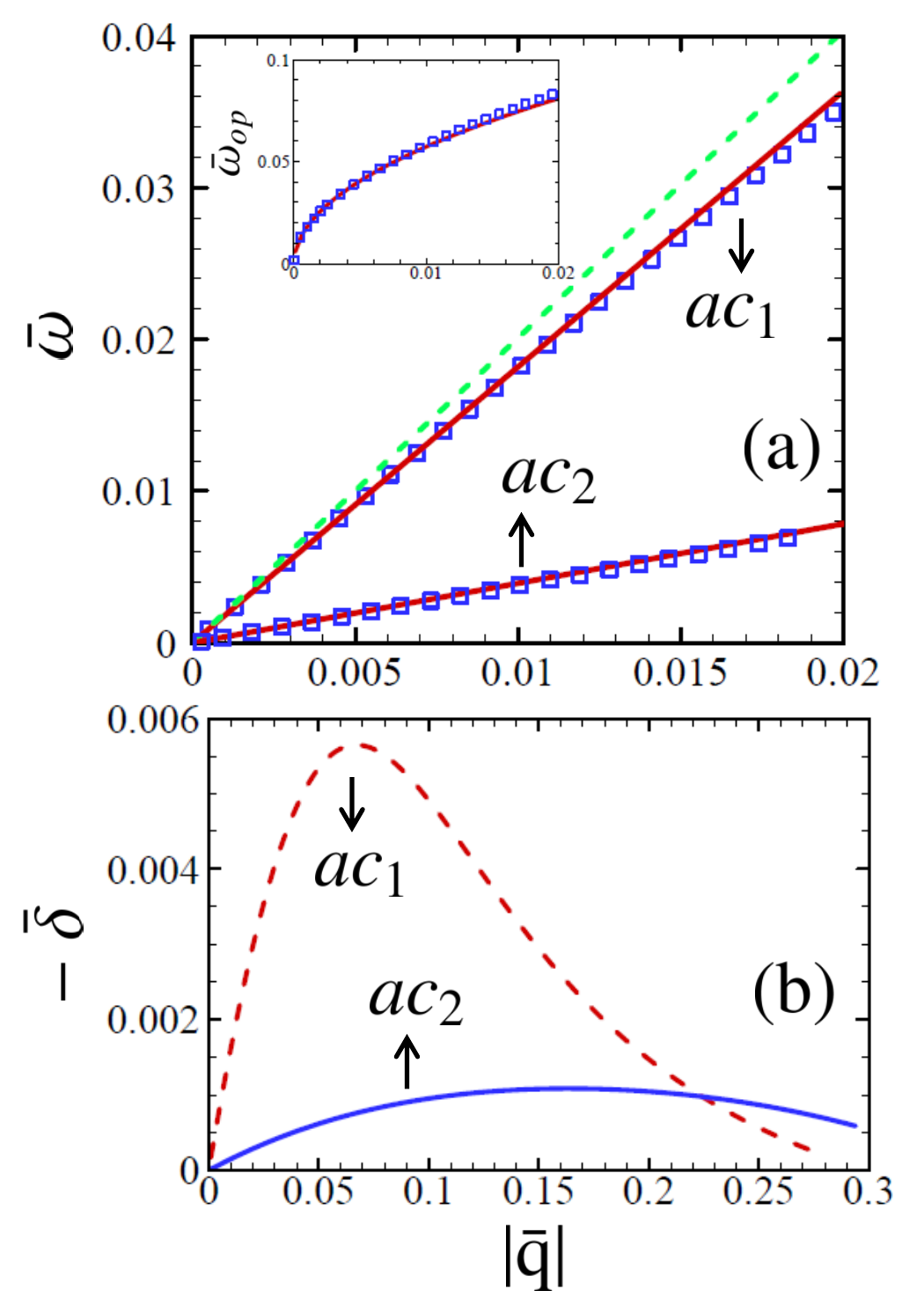}
 \caption{\label{fig1} (Color online) (a)   First (top) and second (bottom) acoustic plasmon dispersion $\bar{\omega}(\bar{q})$ versus $|\mathbf{\bar{q}}|$ for $\bar{\epsilon}=300$. The analytical results (solid lines) of the first acoustic mode are obtained from Eqs.~\eqref{eq:delta} and \eqref{eq:cs1} and for the second acoustic mode we have used Eqs.~\eqref{eq:delta2} and \eqref{eq:cs2}. Numerical results are also shwn for comparison (squares). The dashed green line shows the boundary of the
electron-hole continuum of the first band. Inset: The same comparison for the optical mode of the system, $\bar{\omega}_{op}(\bar{q})$. The long wavelength analytical solid line is obtained from  Eq.~\eqref{eq:w_op1}. (b) The Landau damping of the first and second acoustic modes,  $-\bar{\delta}(\bar{q})$ as a function of $|\mathbf{\bar{q}}|$. }
\end{figure}

\begin{figure}[h]
\centering
\includegraphics[width=1.0\linewidth] {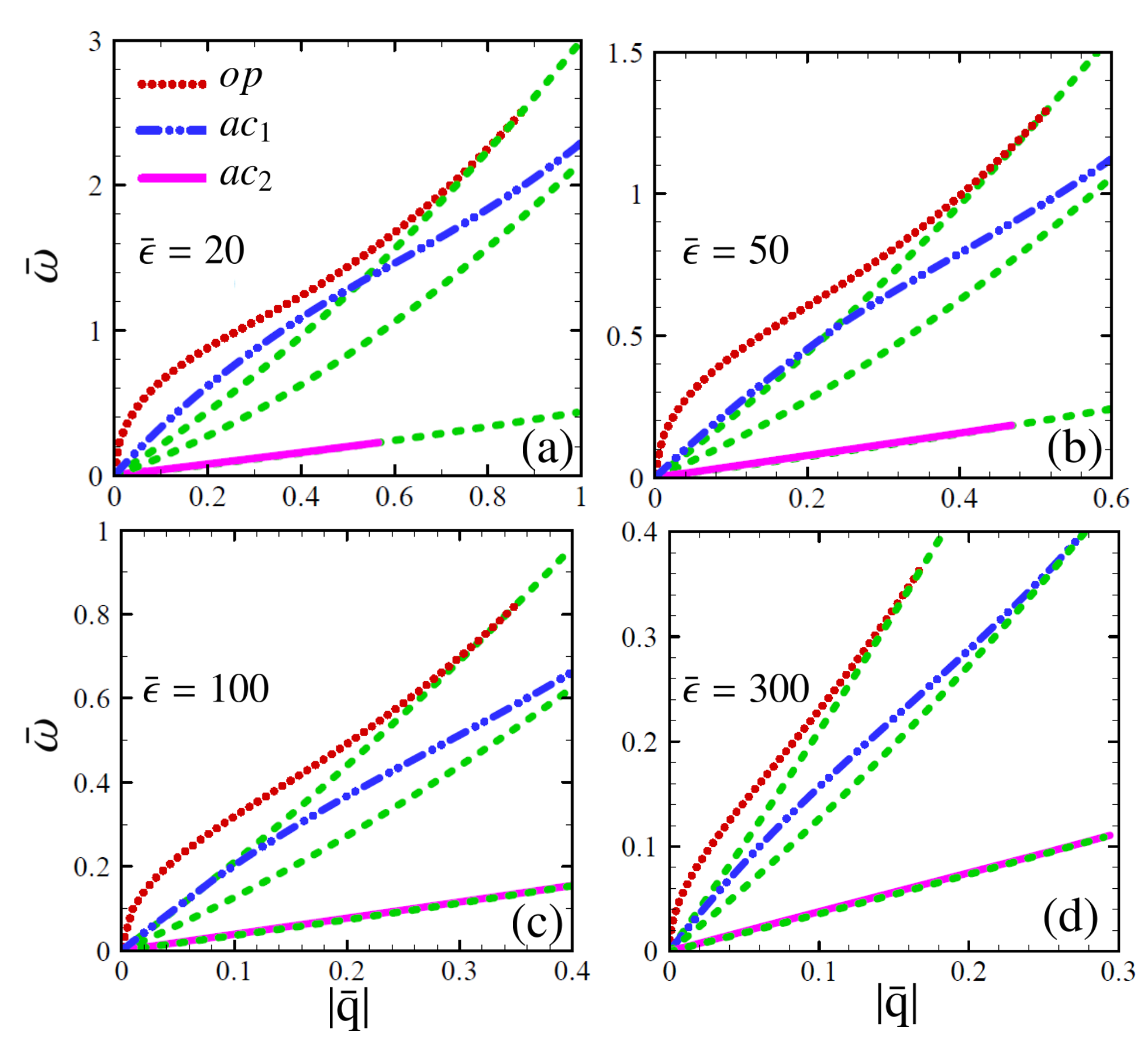}
 \caption{\label{fig:fig2} (Color online) Plasmon modes of the system, $\bar{\omega}(\bar{q})$ for different values of the average dielectric constant as a function of $|\mathbf{\bar{q}}|$ (a) $\bar{\epsilon}=20$, (b) $\bar{\epsilon}=50$, (c) $\bar{\epsilon}=100$ and (d) $\bar{\epsilon}=300$. Dotted red line shows the optical plasmon, the dashed dotted blue line and the solid pink line illustrate the first and second acoustic plasmon modes. The dashed green lines show the boundary of the electron-hole continuum of the bands. For larger values of the dielectric constant of $\rm SrTiO_3$, the critical energy in which the mode becomes damped decreases. }
\end{figure}
\begin{figure}[h]
\centering
\includegraphics[width=1.0\linewidth] {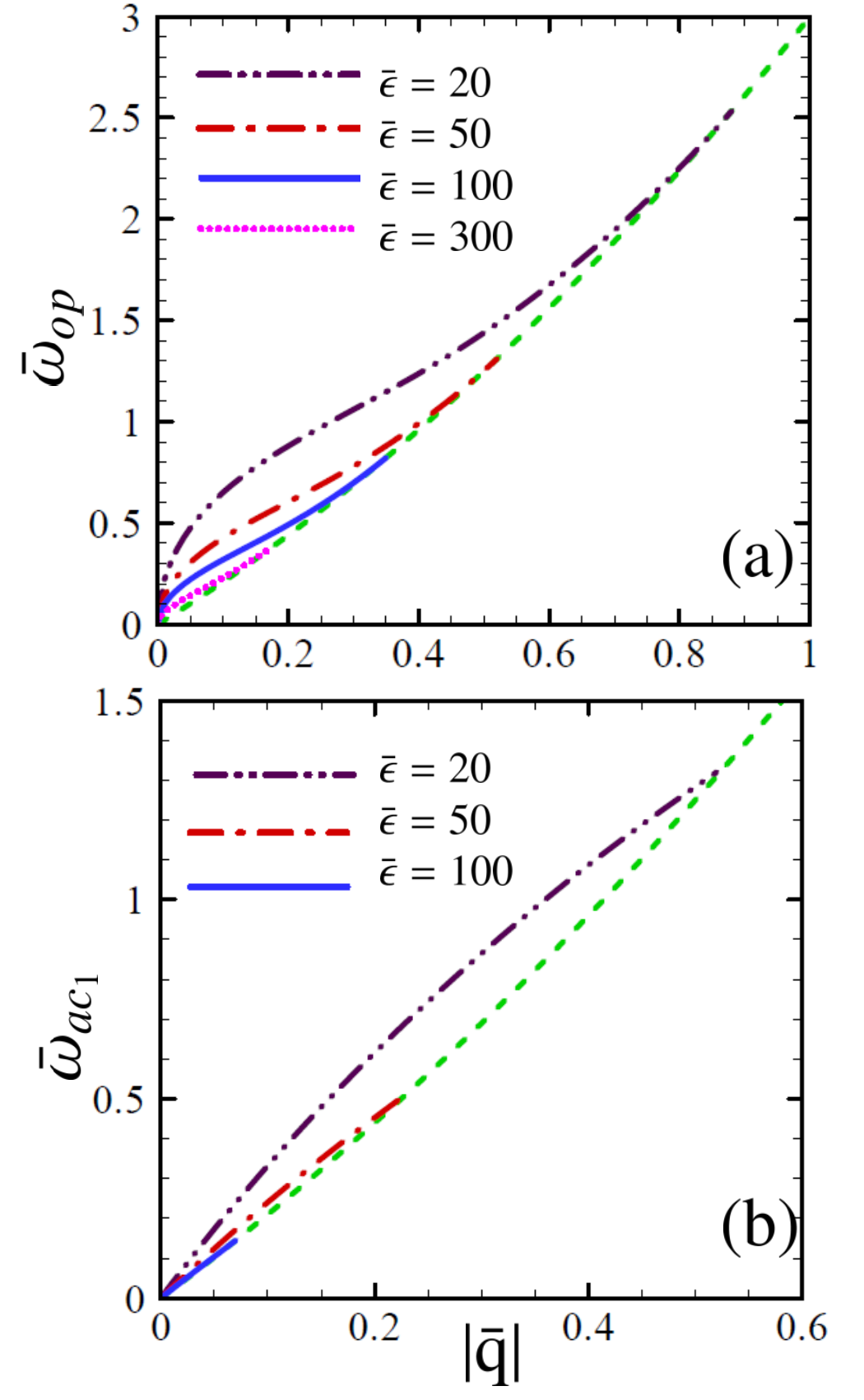}
 \caption{\label{fig:fig3} (Color online) (a) Optical  $\bar{\omega}_{op}(\bar{q})$ and (b) first acoustic mode,  $\bar{\omega}_{ac_1}(\bar{q})$ of 2DEG at interface LAO/STO, for different values of $\bar{\epsilon}$ as a function of $|\mathbf{\bar{q}}|$. For larger dielectric constant of the medium, the system has a shorter Thomas-Fermi wavelength and therefore both optical and acoustic modes touch the particle-hole continuum boundary in a smaller wavelength and become damped. The dashed green lines show the boundary of the
electron-hole continuum of the first band. }
\end{figure}

\subsection{Numerical results and discussions }

In this section we present our numerical findings of plasmon modes at the LAO/STO  interface.

To obtain the collective modes of the interface of LAO/STO, we set $n_{s}=7\times 10^{13}$ cm$^{-2}$, $m_L=0.68 m_e$ and $m_H=7.56m_e$~  \cite{gariglio} where $n_{s}$ is the total electron density of the interface and $m_e$ is the electron mass. The energy gap between $\rm{d_{xy}}$ orbital and other orbitals is $\Delta=50$ meV  and the effective distance between them is assumed to be $a=10 a_0$ with $a_0=3.9$\AA , the lattice constant of $\rm{SrTiO_3}$~  \cite{tolsma}. Here, we use $k_{F,1}$ as a unit of momentum in such a way that $q=k_{{\rm F},1}\bar{q}$ and we use the Fermi energy in the circular band $\varepsilon_{{\rm F},1}=\hbar^2  k_{{\rm F},1}^2/2m_L$ as a unit of energy where $\hbar \omega=\varepsilon_{{\rm F},1}\bar{\omega}$. In all figures, we set $\theta=\pi/2$ otherwise we specify it's value.

To begin with, in Fig. \ref{fig1}(a), the analytical results obtained for optical and acoustic plasmon dispersions in the long wavelength limit $q \rightarrow 0$ are compared with the numerical calculations for $\bar{\epsilon}=300$. The perfect agreement between the analytical results and the numerical calculations of the plasmon modes is confirmed in these figures. The same agreement between analytical and numerical findings holds for other values of dielectric constant as well. In Fig. \ref{fig1}(b), we illustrate our numerical results of the Landau damping of the first and second acoustic plasmon modes of the system as a function of $|{\bar{\rm{q}}}|$ for $\bar{\epsilon}=300$. For the plasmon mode to be physical, the damping should be small in comparison to the plasmon frequency, which can be verified according to Fig. \ref{fig1}(b). The ${\bar \delta}$ tends to zero for larger $|{\bar{\rm{q}}}|$ when the corresponding plasmon modes attain to one of the electron-hole continuum boundaries.

In Fig. \ref{fig:fig2} we show the three branches of collective modes of the system for different values of the dielectric constant of $\rm{SrTiO_3}$, namely $\epsilon_s=15,75,175$ and $575$ for which the average dielectric constant of the system would be $\bar{\epsilon}=20,50,100$ and $300$, respectively (the dielectric constant of $\rm{LaAlO_3}$ is $\epsilon_l=25$). The dotted red line shows the optical plasmon of the system while the dashed dotted blue line and the solid pink line illustrate the first and second acoustic plasmon modes. The dashed green lines show the boundary of the electron-hole continuum of the bands, the region in which $\Im m\chi^0_i(q,\omega)$ differs from zero. The line with the larger slope is the upper boundary of the particle excitations of the band with the fastest carriers below which the imaginary part of the dielectric function has a finite value and the plasmon modes become damped. As seen in this figure, the optical mode emerges always in the undamped region but it becomes damped for higher frequencies. In general, if the quasiparticle goes slightly faster than the
wave, it will be slowed down, and will thus give energy to the collective
mode~\cite{vignale}. On the contrary, if it is slightly slower, it will receive energy
from the collective mode. Under equilibrium conditions, the distribution
of quasiparticles is a decreasing function of their velocity. The net
balance of energy corresponds to an energy transfer from the collective
mode to the individual quasiparticles. In general, once Landau damping becomes possible, the collective mode has such a short lifetime
that it no longer represents a well-defined excitation of the system. Moreover, for larger values of the dielectric constant of $\rm SrTiO_3$, $\epsilon_s$ decreases the critical energy in which this mode becomes damped. The damping of the first and uppermost acoustic mode depends strongly on the value of $\epsilon_s$, while it is undamped at lower energies for lower values of $\epsilon_s$, it becomes completely damped at all energies for high enough values of the dielectric constant. Using Eq.~\eqref{eq:critical}, we can find the critical value for the dielectric constant of the system for the specific value of $a$, above which the first acoustic mode becomes damped at all frequencies, $\bar{\epsilon}_{cr}=164$ which corresponds to $\epsilon_s = 303$.
\begin{figure}[h]
\centering
      \includegraphics[width=1.00\linewidth] {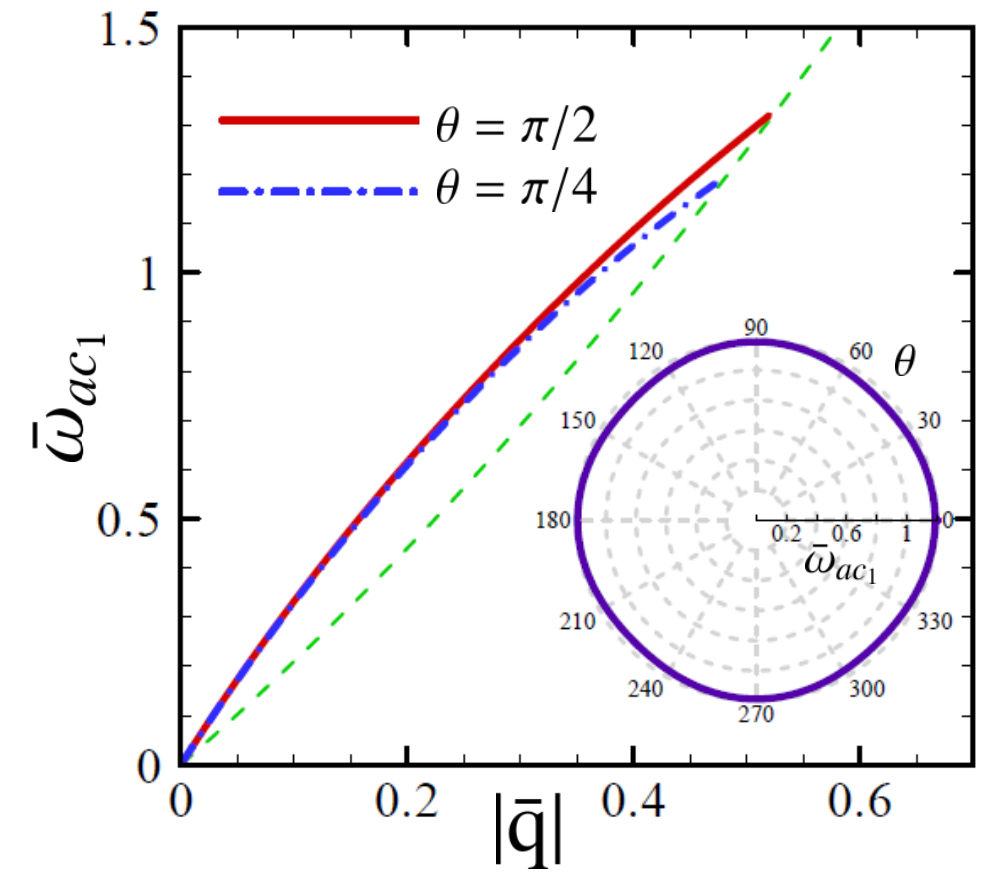}
 \caption{\label{fig4} (Color online) First acoustic plasmon dispersion of the system as a function of $|\mathbf{\bar{q}}|$, $\bar{\omega}_{ac_1}(\bar{q})$ for $\theta=\pi/2$ (solid line) and $\theta=\pi/4$ (dashed dotted line). Inset: Polar plot of the first acoustic mode frequency versus $\theta$, the radial orientation of wave vector for $\bar{q}=0.45$ ($\bar{\epsilon}=20$). The anisotropy of the band dispersion has a small effect on plasmon modes and the modes for wave vectors in different directions are the same, except for larger wave vectors where small deviations appear.  }
\end{figure}
As seen in Fig. \ref{fig:fig2}, for $\bar{\epsilon}>\bar{\epsilon}_{cr}$ (Fig. \ref{fig:fig2}(d)) the first acoustic plasmon is entirely damped, but for  $\bar{\epsilon}<\bar{\epsilon}_{cr}$, we can find a region in the limit $q\to0$ where the first acoustic mode is undamped. In the case $\bar{\epsilon}=300$, the critical effective distance $a$ above which the undamped first acoustic plasmon emerges is $7.1\times 10^{-7}$cm$^{-2}$ that is almost twice larger than the effective distance we employed. The second acoustic plasmon mode, as we mentioned before, is strongly damped in all cases.
\begin{figure}[h]
\centering
      \includegraphics[width=1.0\linewidth] {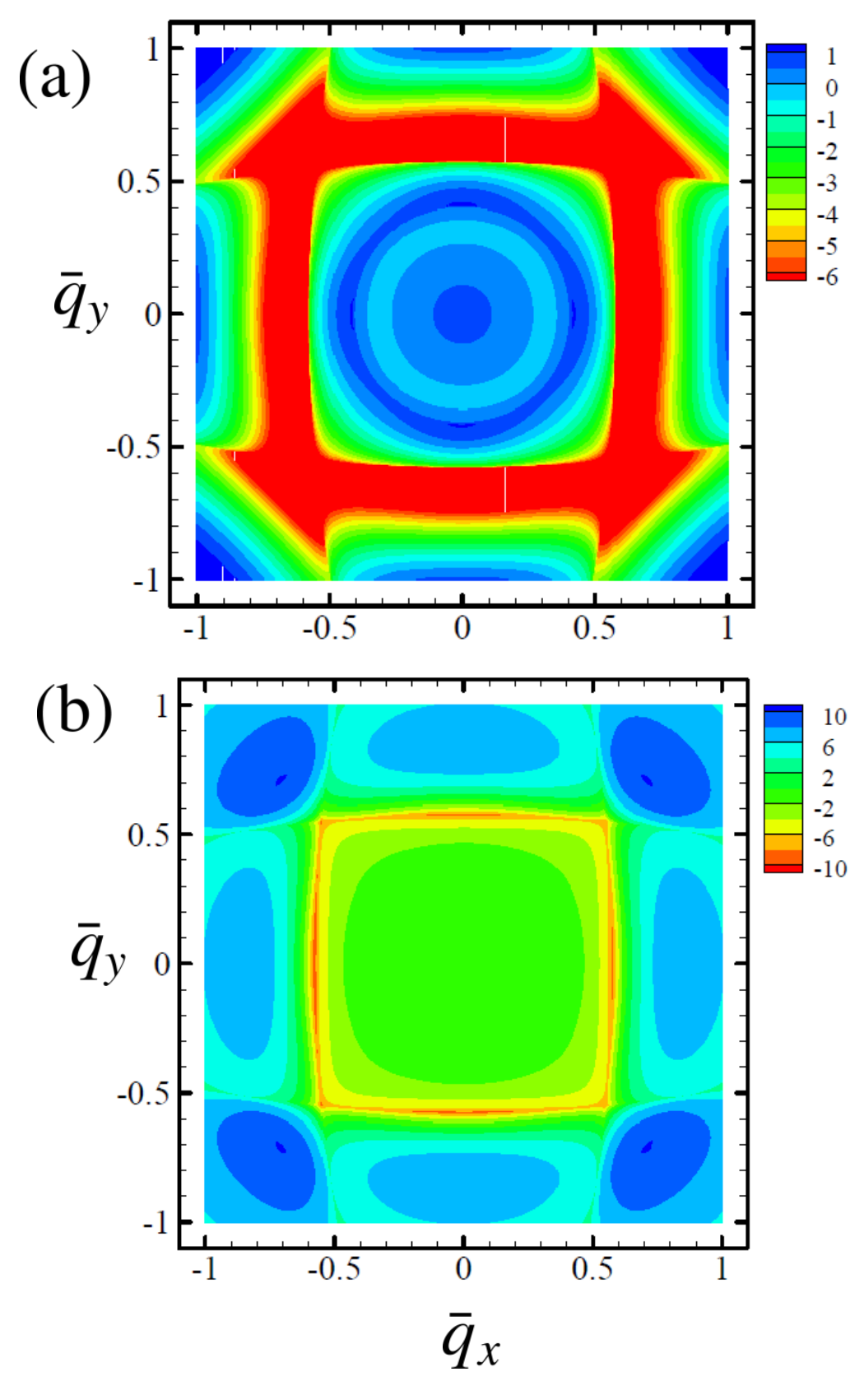}
 \caption{\label{fig5} (Color online) Cross section of (a) real and (b) imaginary parts of the RPA dielectric function $\varepsilon(\bar{\rm{q}},\bar{\omega})$  Eq.~\eqref{eq:eps} versus $\bar{q}_x$ and $\bar{q}_y$ for $\bar{\omega}=1$ ($\bar{\epsilon}=20$). The dielectric function becomes more isotropic for smaller wave vectors. Therefore, the optical mode is completely isotropic and the acoustic mode is only weakly anisotropic. }
\end{figure}

In Fig. \ref{fig:fig3}(a), we compare the optical modes of the system for different values of $\bar{\epsilon}$. In Fig. \ref{fig:fig3}(b) the same comparison is made for the first acoustic mode. The damped part of the  modes is not illustrated in this figure. For larger dielectric constant of the medium,  the system has a shorter Thomas-Fermi wavelength and therefore both optical and acoustic modes touch the particle-hole continuum boundary in a smaller wavelength and become damped. Having mentioned before, for $\bar{\epsilon}=300$, the first acoustic mode is entirely damped, so it is not shown in Fig. \ref{fig:fig3}(b).

Since the band dispersion of LAO/STO is anisotropic, the response function is a function of both the absolute value of the wave vector $q$  and it's angle $\theta$.
To compare the results of the plasmon modes for different orientations of the wave vector $\mathbf q$, in Fig. \ref{fig4} we show the first acoustic plasmon for two different angles $\theta=\pi/2$ and $\theta=\pi/4$. In the inset, once again we see the first acoustic mode frequency of the system as a function of $\theta$ for $\bar{q}=0.45$. The figures are plotted for $\bar{\epsilon}=20$ and the
dashed green lines show the boundary of the electron-hole continuum for $\theta=\pi/2$. The anisotropy of the band dispersion has a small effect on plasmon modes and the modes for wave vectors in different direction are exactly the same, except for larger wave vectors where small deviations appear. It is interesting to note that the optical mode is the same for all angles of $\mathbf q$. In the long wavelength limit, this can be approved analytically considering Eq.~\eqref{eq:w_op1}.

To better understand this almost unexpected behavior, in Fig. \ref{fig5}, we plot the real and imaginary parts of the dielectric function of the system as a function of $\bar{q}_x$ and $\bar{q}_y$ for $\bar{\omega}=1$. It is clear from the figures that the dielectric function of the system becomes more and more isotropic as we move into smaller wave vectors. Therefore, the optical mode, which for a specified frequency occurs at the smaller wave vector, is completely isotropic and the acoustic mode is only weakly anisotropic. On the other hand, for all the values of electron density we can assume, the fastest carriers of the system are those of circular band $\rm{d_{xy}}$. These electrons form an isotropic electron-hole continuum boundary ($\omega=v_{F1}q$) above which the undamped plasmon modes appear. In order to see more explicitly the effect of the anisotropic bands on plasmon modes, the slope of the electron-hole continuum boundary of at least one of the elliptical bands should be larger than that of the circular band. In this case the boundary of the electron-hole continuum of the fastest carriers become anisotropic (Eqs.~\eqref{eq:v2} and \eqref{eq:v3}) and a more anisotropic behavior for the collective modes is expected. But for this condition to be fulfilled, at least we should have $n_2>\frac{m_H}{m_L}n_1$, the condition is never satisfied for any value of the band offset.
\begin{figure}[h]
\centering
      \includegraphics[width=1.\linewidth] {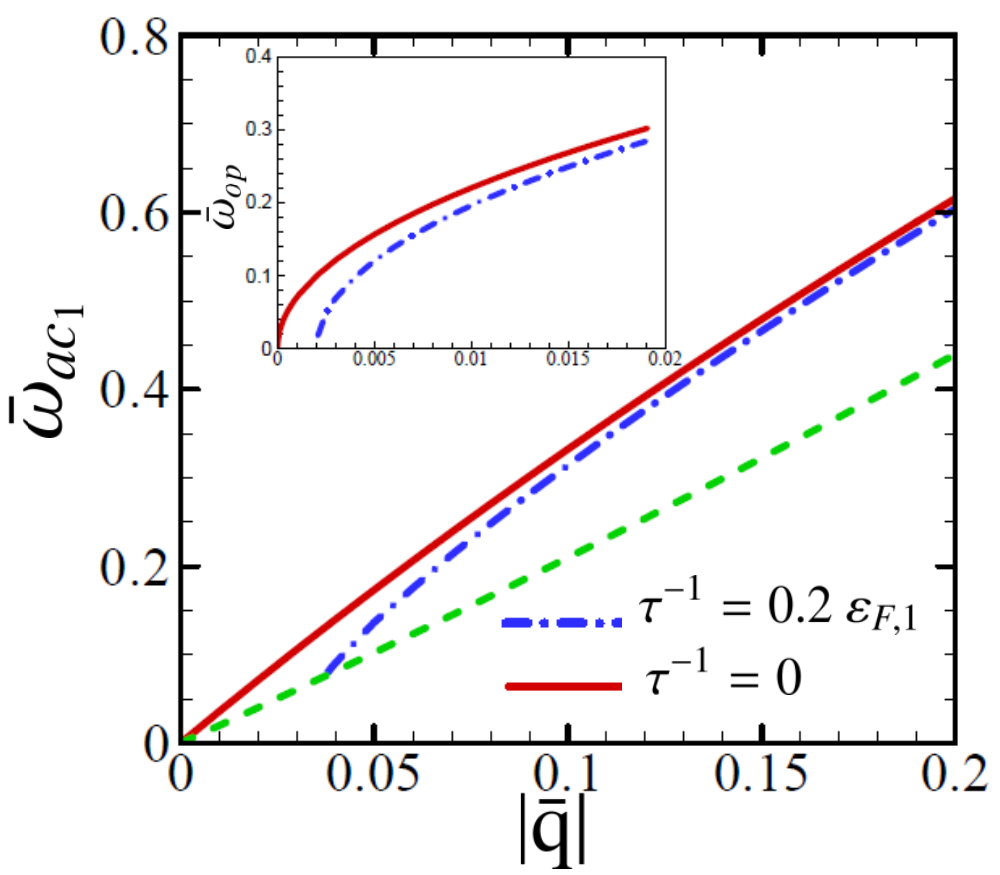}
 \caption{\label{fig6} (Color online) First acoustic plasmon dispersion, $\bar{\omega}_{ac_1}(\bar{q})$ as a function of $|\mathbf{\bar{q}}|$ in the presence of disorder (dashed dotted line) for $\tau^{-1}=0.2~\varepsilon_{F,1}$. The plasmon mode for the pure system is also shown for comparison (solid line). The dashed green line is the boundary of the electron-hole continuum and the results are obtained for $\bar{\epsilon}=20$. Inset: Optical plamon mode, $\bar{\omega}_{op}(\bar{q})$ in the presence of disorder (dashed dotted line) and for a pure system (solid line). }
\end{figure}

In order to understand the impact of weak disorder on the plasmon modes, we modify slightly the noninteracting response function to capture that effect at the long wavelength limit. To do so, the noninteracting density-density response functions in Eq.~\eqref{eq:eps} should be replaced by the disorder averaged response functions of each band~\cite{vignale} which for the isotropic 2DEG has the following form
\begin{equation}\label{eq:imp}
\chi^0_{imp}(\mathbf{q,\omega})\approx -N(0)\big(1-\frac{\omega}{\sqrt{(\omega+i/\tau)^2-v_F^2q^2}-i/\tau}\big)
\end{equation}
where $\tau$ is the elastic lifetime of the momentum eigenstates in a disordered system and $N(0)=m/\pi \hbar^2$ with $m$ being the effective mass. Note that for the elliptical bands, transformations similar to Eqs.~\eqref{eq:chi02} and \eqref{eq:chi03} should be applied. In Fig. \ref{fig6} we illustrate the effect of weak disorder on the plasmon modes of the system. It is clear from the figure that the influence of disorder on the acoustic mode is considerable and this mode disappears completely for $|\mathbf{q}|<0.04$ at the given density. The same effect can be seen for the optical mode in the inset, but in this case the plasmon mode vanishes for a longer wavelength $|\mathbf{q}|<0.002$. In both cases, disorder also decreases the plasmon energy. Consequently, in the presence of disorder it is difficult to access plasmon modes experimentally especially at the long wavelength limit. It is also worthwhile to note that in the low electron densities and hence the large dielectric constant of the medium, the modes are closer to the electron-hole boundary and therefore it would be even more difficult to observe the plasmon modes in the presence of disorder in the system. The precise influence of defects and impurities on the plasmon modes has not been explored and it remains an open question in the system.

\section{Plasmon modes of graphene-LAO/STO double layer}

\begin{figure}[h]
\centering
      \includegraphics[width=1.\linewidth] {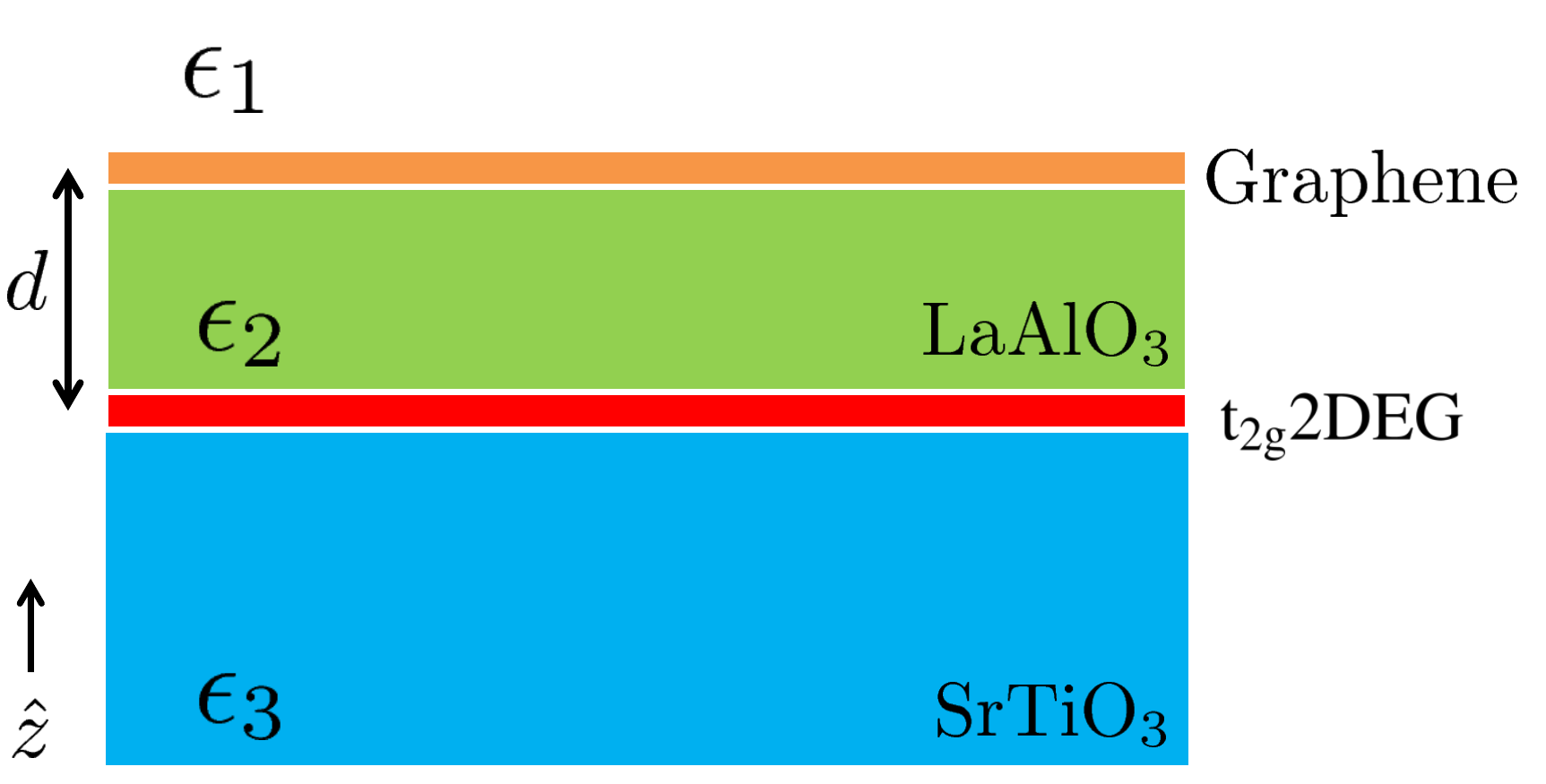}
 \caption{\label{fig7} (Color online)  A schematic of the double layer system constructed from graphene at the top layer and a two-dimensional $\rm{t_{2g}}$ electron gas at the interface of $\rm{LaAlO_3}$ and $\rm{SrTiO_3}$ as the bottom layer.   }
\end{figure}

In this Section, we would like to consider a structure combining a doped graphene layer deposited on the top of $\rm{LaAlO_3}$. Therefore, we have a double layer structure with graphene on the top and a three-band 2DEG at the interface of LAO/STO, separated from each other by a distance $d$ which in this case it would be the thickness of $\rm{LaAlO_3}$ with dielectric constant $\epsilon_2$ as shown in Fig.~\ref{fig7}. Assuming the subsystems are unhybridized, the interlayer tunneling can be neglected and Coulomb interaction is the only source of coupling between layers. The Hamiltonian of the system then reads
\begin{equation}
 \begin{split}\label{eq:fullH}
{\hat {\cal H}} &= \hbar v_{\rm D}\sum_{{\bm k}, \alpha, \beta} {\hat \psi}^\dagger_{{\bm k}, \alpha}
( {\bm \sigma}_{\alpha\beta} \cdot {\bm k} ) {\hat \psi}_{{\bm k}, \beta}
+
\frac{1}{2 S}\sum_{{\bm q} \neq {\bm 0}} V_{ gg}(q) {\hat \rho}^{({ g})}_{\bm q} {\hat \rho}^{({ g})}_{-{\bm q}}\\
&+ \sum_{ k,\theta, i}( \frac{\hbar^2 { k}^2\cos^2(\theta)}{2m_{ x,i}}+\frac{\hbar^2 { k}^2\sin^2(\theta)}{2m_{ y,i}}+\Delta_{ i}){\hat \phi}^\dagger_{\bm k, i}{\hat \phi}_{\bm k, i}\\
&+\frac{1}{2 S}\sum_{{\bm q} \neq {\bm 0}, i, j} V_{ ij}(q) {\hat \rho}^{({ i})}_{\bm q} {\hat \rho}^{({ j})}_{-{\bm q}}\\
&+ \frac{1}{2 S}\sum_{{\bm q}, i} V_{ gi}(q) ({\hat \rho}^{({ g})}_{\bm q} {\hat \rho}^{({ i})}_{-{\bm q}}+{\hat \rho}^{({ i})}_{\bm q} {\hat \rho}^{({ g})}_{-{\bm q}})
\end{split}
\end{equation}
where $v_{\rm D}\approx10^6 \rm m/s$ is the Dirac velocity, $\bm \sigma$ is a vector whose components are the Pauli matrices $\sigma^x$ and $\sigma^y$,  $\alpha$ and $\beta$ are sublattice indices in graphene, ${\hat \psi}^\dagger_{{\bm k}, \alpha}$ and ${\hat \psi}_{{\bm k}, \alpha}$ are creation and annihilation operators in graphene, $S$ is the sample area, $V_{ gg}(q)$ is the Fourier transform of the bare Coulomb interaction between electrons in graphene and ${\hat \rho}^{({ g})}_{\bm q} = \sum_{{\bm k}, \alpha} {\hat \psi}^\dagger_{{\bm k} - {\bm q}, \alpha}{\hat \psi}_{{\bm k}, \alpha}$ is the density operator in graphene. The terms in the second and third lines are the Hamiltonian of the 2DEG at the interface described in previous section and finally $V_{ gi}(q)$  in the last line is the Coulomb interaction between electrons in graphene and the electrons of band $i$ at the interface. In this case, since the distance between layers, $d$, is a few nm we can neglect $a$, the effective distance between $\rm{t_{2g}}$ orbitals that we assumed in the previous section.

Following some simple and straightforward electrostatic calculations~\cite{profumo}, it has been shown that for multilayer structures the Coulomb interaction is changed such that in our case we have
\begin{equation}\label{eq:v11}
V_{ gg}(q) = \frac{4\pi e^2}{q D(q)} [ (\epsilon_2 + \epsilon_3) e^{qd} +  (\epsilon_2 - \epsilon_3) e^{-qd}]
\end{equation}
and
\begin{equation}\label{eq:v12}
V_{ gi}(q) = V_{ ig}(q) = \frac{8\pi e^2}{q D(q)}~\epsilon_2
\end{equation}
with
\begin{equation}
D(q) =  [(\epsilon_1 + \epsilon_2)(\epsilon_2 + \epsilon_3) e^{qd}+(\epsilon_1 - \epsilon_2)(\epsilon_2 - \epsilon_3) e^{-qd}]
\end{equation}
and $V_{ ij}(q)=V_{ gg}(q)\vert_{\epsilon_3\leftrightarrow \epsilon_1}$. Note that as we neglect the effective distance between circular and elliptical bands here, $V_{ ij}$ is the same for all values of $ i$ and $ j$. To find the collective modes of the system within RPA, once more we should employ Eq.~\eqref{eq:xi}. In this case $\boldsymbol{\hat{\chi}^0}(\mathbf{q,\omega})$ is the $4\times 4$ diagonal noninteracting density-density response matrix of the system with elements ${\chi^0_g}(q,\omega)$, the noninteracting density-density response function of graphene~\cite{e.h.hwang}, and ${\chi^0_1}(q,\omega)$, ${\chi^0_2}(\mathbf{q,\omega})$ and ${\chi^0_3}(\mathbf{q,\omega})$ which we have introduced before. Inverting Eq.\eqref{eq:xi} the dielectric function of the system reads
\begin{equation}\label{eq:ep2}
 \begin{split}
\varepsilon(\mathbf{q,\omega})& = [1-V_{ gg}(q) \chi_{ g}^0(q,\omega)]\\
&\times[1-V_{ ij}(q)(\chi_{1}^0(q,\omega)+\chi_2(\mathbf{q,\omega})+\chi_3(\mathbf{q,\omega}))] \\
&- V^2_{ gi}(q)\chi_{ g}^0(q,\omega)[\chi_{1}^0(q,\omega)+\chi_{2}^0(\mathbf{q,\omega}+\chi_{3}^0(\mathbf{q,\omega}]
 \end{split}
\end{equation}

The zeros of the dielectric function are the collective modes of the system. Here again we can find an optical higher frequency plasmon mode ($\omega_{op}(q\to0)\propto\sqrt{q}$) which can be found in both monolayer and double layer systems and as discussed before, emerges from the in-phase oscillations of the electrons of two layers. In double layer systems, there is also a lower frequency acoustic mode ($\omega_{ac}(q\to0)\propto q$) whose occurrence above the particle-hole continuum of the system depends on characteristics of the system such as the distance between two layers and the carrier density of each layer~\cite{santoro,dassarma1,rosario,principi,stauber}. In the system under consideration, since we have a three-band electron gas at the bottom layer, we expect to have three acoustic branches, two of which are located below the electron-hole continuum and hard to detect but as the previous case, the uppermost acoustic mode can emerge above the boundary of the single particle excitation region.

\subsection{Analytical results at long wavelength limit}

At the long wavelength limit $q\to0$, we can find an analytical expression for both optical and acoustic collective modes of the system. In this region and for $\omega \gg qv_D$  (since for all acceptable density values $v_D$ is larger than the largest Fermi velocity of $ \rm{t_{2g}}$ bands), we can again make use of Eq.~\eqref{eq:chi0app} for the density-density response functions of the 2DEG at the interface and also the limiting form of the density-density response function of graphene~\cite{e.h.hwang}
\begin{equation}
 \chi^0_{g}(q,\omega)\simeq \sqrt{\frac{N_gn_g}{\pi}}\frac{v_Dq^2}{2\hbar\omega^2}[1-(\frac{\omega}{2v_Dk_{F,g}})^2]
\end{equation}
with $n_g$ the electron density in graphene layer, $N_g=4$ (spin and valley flavors in graphene layer) and the Fermi wave vector of graphene defined as $k_{{\rm F},g}=\sqrt{4\pi n_g/N_g}$. Substituting these expressions in Eq.\eqref{eq:ep2}, the long wavelength limit of the optical plasmon mode of the system leads to
\begin{equation}\label{eq:w_opg}
\omega_{op}^2(q\to0)=\frac{2\pi e^2}{\bar{\epsilon} }\Bigl[\sqrt{\frac{N_gn_g}{\pi}}\frac{v_D}{2\hbar}+\frac{n_1}{m_L}+n_2( \frac{1}{m_L}+\frac{1}{m_H})\Bigr] q
\end{equation}
with $\bar{\epsilon}=(\epsilon_1+\epsilon_3)/2$ ($\epsilon_1$ and $\epsilon_3$ are the top and bottom dielectric constants). Note that similar to other double layer systems, the optical plasmon mode is independent of the dielectric material between layers and it is found as the combination of plasmon modes of each subsystem individually.

To find the group velocity of the acoustic plasmon lying above the electron-hole continuum, we follow the same procedure described in the previous section. After some algebra we end up with the following equation for $x=c_s/v_D$
\begin{equation}\label{eq:csg}
\begin{split}
&R(x)[\epsilon_2\sqrt{x^2-1}+2N_gdk_{F,g}\alpha_{ee}f(x)]\\
&+2N_g\epsilon_2\alpha_{ee}h_1(x)h_2(x)h_3(x)f(x)=0
\end{split}
\end{equation}
where
\begin{equation}\label{eq:r}
\begin{split}
&R(x)=\\
&\Gamma_1g_1(x)h_2(x)h_3(x)+\Gamma_2h_1(x)[g_2(x)h_3(x)+g_3(x)h_2(x)]
\end{split}
\end{equation}
and $\Gamma_i=2N_s/(a_{B,i}k_{F,g})$ (with $N_s=2$ the spin flavor in 2DEG, $a_{B,1}=\hbar^2/(m_Le^2)$ and $a_{B,2}=\hbar^2/(m_De^2)$), $h_i(x)=\sqrt{x^2N_sr_{s,i}^2-4\alpha_{ee}^2}$, $g_i(x)=h_i(x)-x\sqrt{N_s}r_{s,i}$ , $f(x)=\sqrt{x^2-1}-x$, $\alpha_{ee}=e^2/\hbar v_D$ and we define $r_{s,i}=1/(a_{B,i}\sqrt{\pi n_i})$ with $n_i$ the electron density of band $i$ of 2DEG at the interface.

$c_s=v_D$ or $x=1$ is the minimum value of the sound velocity above which the acoustic plasmon is undamped at the long wavelength limit. In this case we define a critical value for the distance between layers $d_{cr}$ as
\begin{equation}\label{eq:dcritical}
d_{cr}=-\frac{\epsilon_2h_1(1)h_2(1)h_3(1)}{k_{F,g}R(1)}
\end{equation}

When the distance between layers is larger than $d_{cr}$, the acoustic mode is lying out of the electron-hole continuum and is undamped unless it touches the boundary of the continuum, but for distances less than $d_{cr}$ the acoustic mode is completely damped. That is because for smaller distances the out-of-phase oscillations of the electrons is not easy owing to the interlayer Coulomb interaction. The expressions \eqref{eq:csg}-\eqref{eq:dcritical} are also generalization of the results of Ref. [\onlinecite{principi}] (obtained for graphene-2DEG double layer) for multiband and anisotropic 2DEG.
\begin{figure}[h]
\centering
      \includegraphics[width=1.\linewidth] {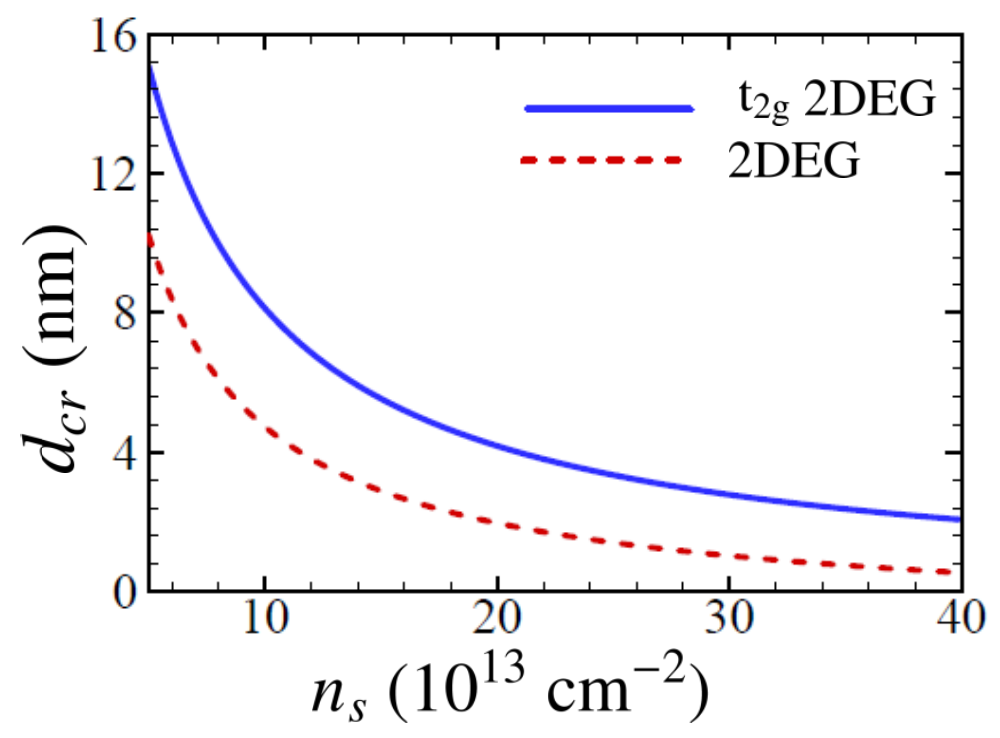}
\caption{\label{fig8} (Color online)  Variation of interlayer critical separation, $d_{cr}$ with electron density, $n_s$ at the interface for $\rm{t_{2g}}$ electron gas (solid line) and isotropic one band 2DEG (dashed line). The acoustic plasmon of the isotropic one-band 2DEG emerges in a smaller interlayer separation. This can be owing to the stronger Coulomb coupling between layers when we have $\rm{t_{2g}}$ electron gas at the interface. Thus, the acoustic mode begins to appear in a larger interlayer separation. }
\end{figure}

In Fig.~\ref{fig8} we show how $d_{cr}$ varies with increasing the electron density at the interface. We also illustrate the same results when an isotropic one band 2DEG with $m^*=m_L$ replaces t$_{2g}$ 2DEG. This can be the case when the elliptical bands at the interface are not populated. It can be seen that the critical value of the interlayer distance decreases as we increase the electron density of the interface but it is independent of the density of graphene (the latter can be concluded from Eqs.\eqref{eq:r} and \eqref{eq:dcritical}). The reason is that the interaction of 2DEG and hence $d_{cr}$ decreases as the density increases, but the interaction parameter ($\alpha_{ee}$) of graphene is independent of the density. Comparing the case of 2DEG and that of $\rm{t_{2g}}$, we can see that for the whole range of electron density at the bottom layer, the acoustic plasmon of the isotropic 2DEG emerges in a smaller interlayer separation. This can be owing to the stronger Coulomb coupling between layers when we have $\rm{t_{2g}}$ electron gas at the interface, so that the acoustic mode begins to appear in a larger interlayer separation.

We can also see that similar to other double layer structures, the emergence of an acoustic mode only depends on the dielectric material between two layers ($\rm{LaAlO_3}$ in this case). Note that although emerging the acoustic mode at the long wavelength limit depends only on $\epsilon_2$, the dielectric constant of $\rm{SrTiO_3}$ still has a role, as for smaller values of $\epsilon_3$ the damping of the acoustic plasmon mode occurs in larger wave vectors.

\subsection{Numerical results and discussion }
\begin{figure}[h]
\centering
      \includegraphics[width=1.\linewidth] {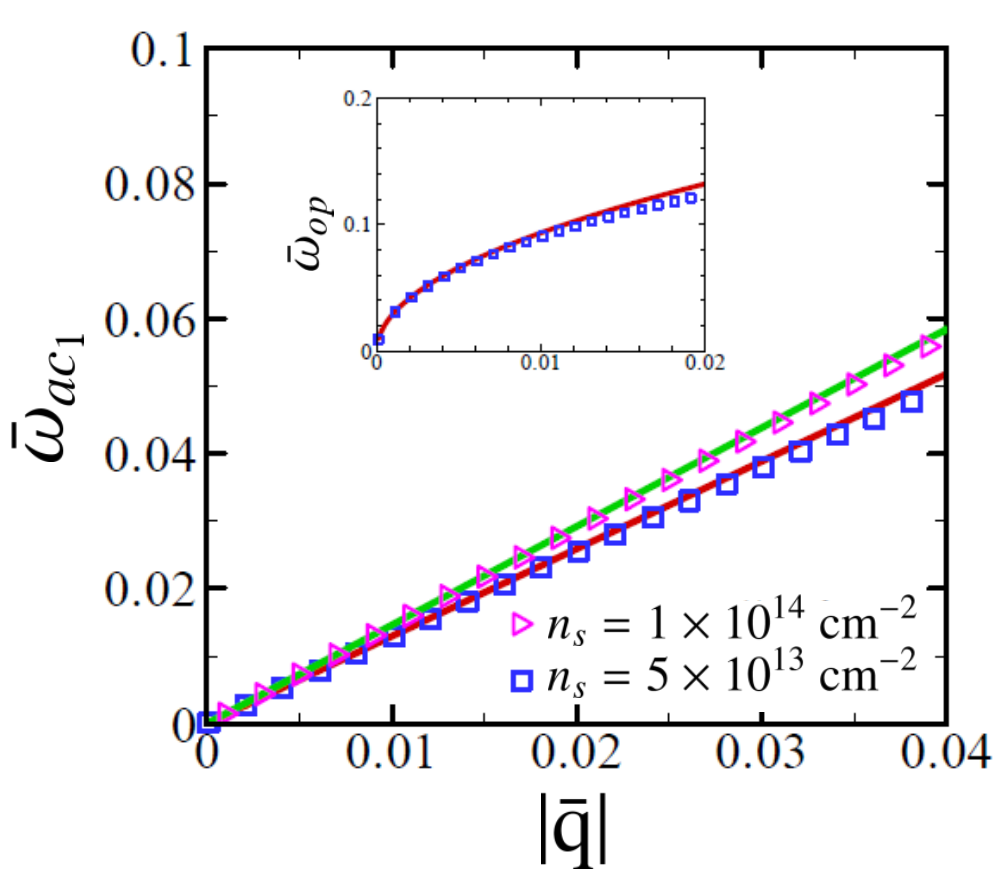}
 \caption{\label{fig9} (Color online) First acoustic plasmon dispersion, $\bar{\omega}_{ac_1}(\bar{q})$ versus $|\bar{\mathbf{q}}|$ for two values of electron density at the interface. Numerical results(squares and triangles) are obtained solving  Eq.~\eqref{eq:ep2} and long wavelength analytical results (solid lines)  are obtained from  Eq.~\eqref{eq:csg}).   Inset: Optical mode  dispersion of the system, $\bar{\omega}_{op}(\bar{q})$ for $n_s=\rm{5\times 10^{13}~cm^{-2}}$. The long wavelength analytical solid line is obtained from  Eq.~\eqref{eq:w_opg}. }
\end{figure}
\begin{figure}[h]
\centering
      \includegraphics[width=1.\linewidth] {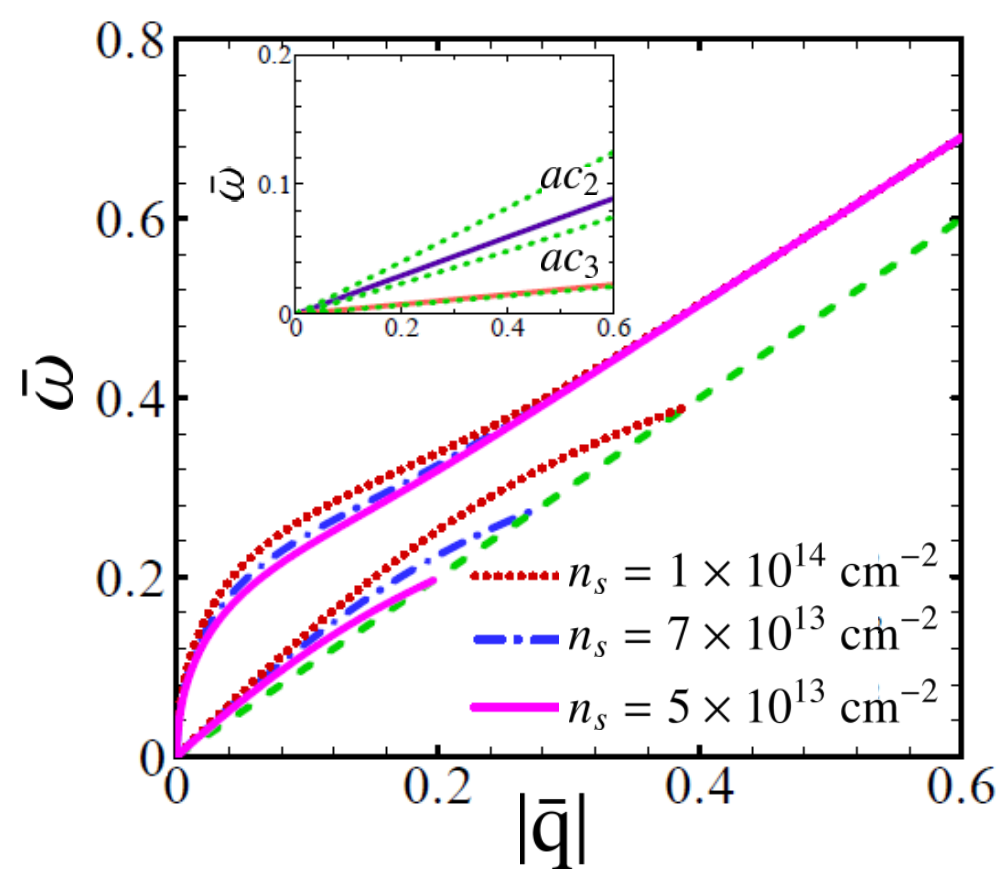}
 \caption{\label{fig10} (Color online) Optical plasmon (upper graphs) and first acoustic plasmon  (lower graphs) dispersions versus $|\bar{\mathbf{q}}|$ for three values of electron density at the interface. The dashed green line is also the boundary of the electron-hole continuum of graphene. Inset: Two solid lines in this figure belong to the second and third damped acoustic modes for $n_s=\rm{7\times 10^{13}~cm^{-2}}$. Here again the dashed green lines refer to the upper boundary of electron-hole continuum of each \rm{d} band.  }
\end{figure}
In this part we find the behavior of the collective modes of the system in the whole $(q,\omega)$ plane. All figures are plotted assuming $n_g=\rm{10^{12}~cm^{-2}}$, $\epsilon_1=1$ for the air, $\epsilon_2=25$ for $\rm LaAlO_3$ and $\epsilon_3=20$ for $\rm SrTiO_3$ and $d=40$ nm. Note that in this section we use $k_{{\rm F},g}$ as unit of the momentum where $q=k_{{\rm F},g}\bar{q}$ and Fermi energy in the graphene layer $\varepsilon_{{\rm F},g}=\hbar v_D k_{{\rm F},g}$ as unit of energy where $\hbar \omega=\varepsilon_{{\rm F},g}\bar{\omega}$. In all figures we consider $\theta=\pi/2$.

In Fig. \ref{fig9} we illustrate the acoustic mode of the system at the long wavelength limit. We plot the acoustic mode for two values of interface densities $n_s=\rm{5\times10^{13}~cm^{-2}}$ and $n_s=1\rm{\times10^{14}~cm^{-2}}$. The squares and triangles are acoustic modes of the system obtained solving Eq.~\eqref{eq:ep2} numerically and the solid lines show the analytical result obtained from Eq.~\eqref{eq:csg} for long wavelength limit. In the inset, we compare the long wavelength analytical optical plasmon of the system (obtained from Eq.~\eqref{eq:w_opg}) with numerical results. Note that as the analytical expressions are valid for $qd\ll 1$, for smaller interlayer separations the agreement between analytical and numerical results holds in larger wave vectors.
\begin{figure}[h]
\centering
      \includegraphics[width=0.95\linewidth] {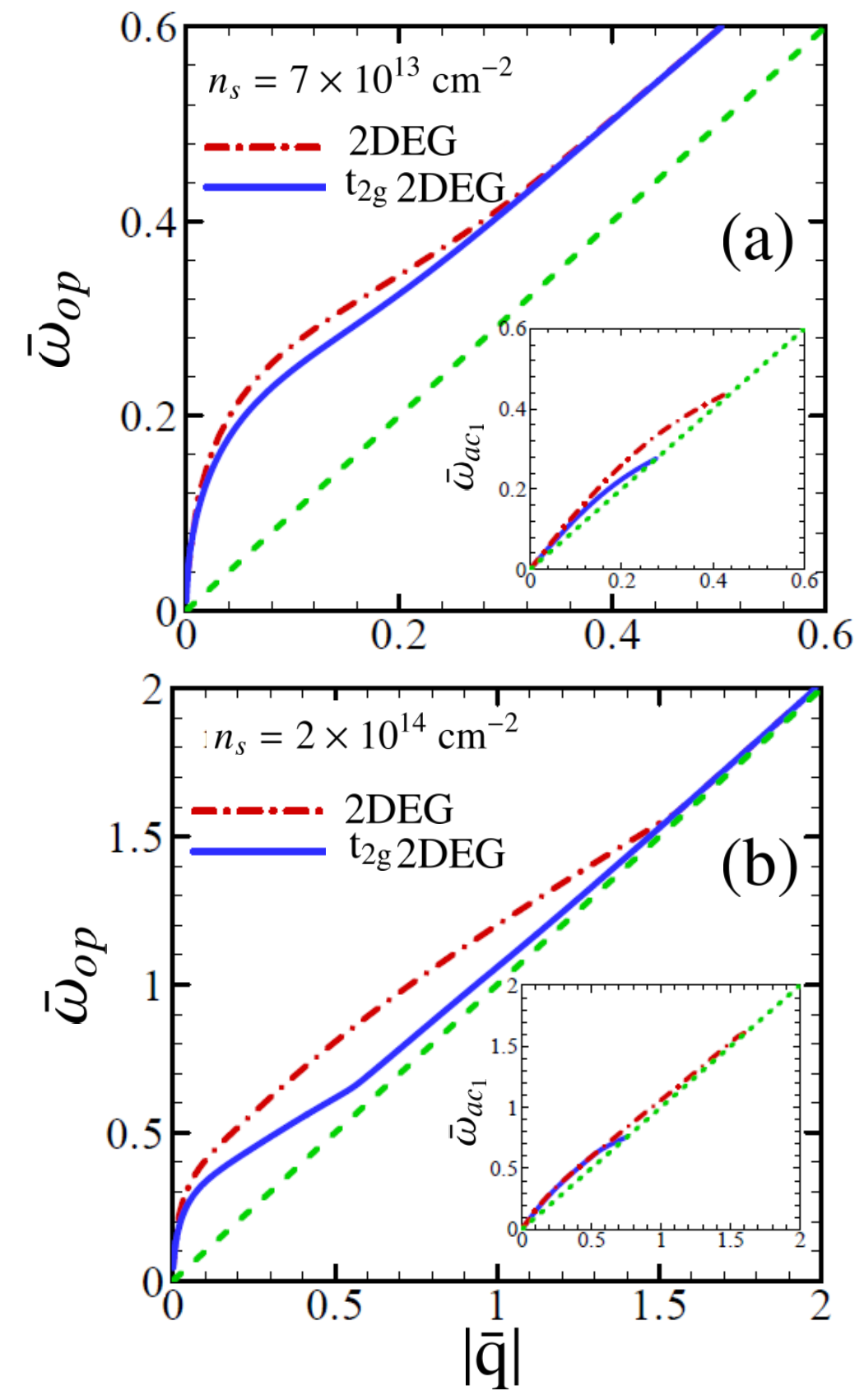}
 \caption{\label{fig11} (Color online) (a) Optical mode dispersion  of the system,  $\bar{\omega}_{op}(\bar{q})$ versus $|\bar{\mathbf{q}}|$ with the electron density $n_s=\rm{7\times 10^{13}~cm^{-2}}$ at the interface(solid line) in comparison with the optical mode of a system with isotropic 2DEG at the interface with the same electron density(dashed dotted line). Inset: Comparison between acoustic modes of the systems,  $\bar{\omega}_{ac_1}(\bar{q})$. (b) Same as in panel (a) but for $n_s=\rm{2\times 10^{14}~cm^{-2}}$.   }
\end{figure}

\begin{figure}[h]
\centering
      \includegraphics[width=1.\linewidth] {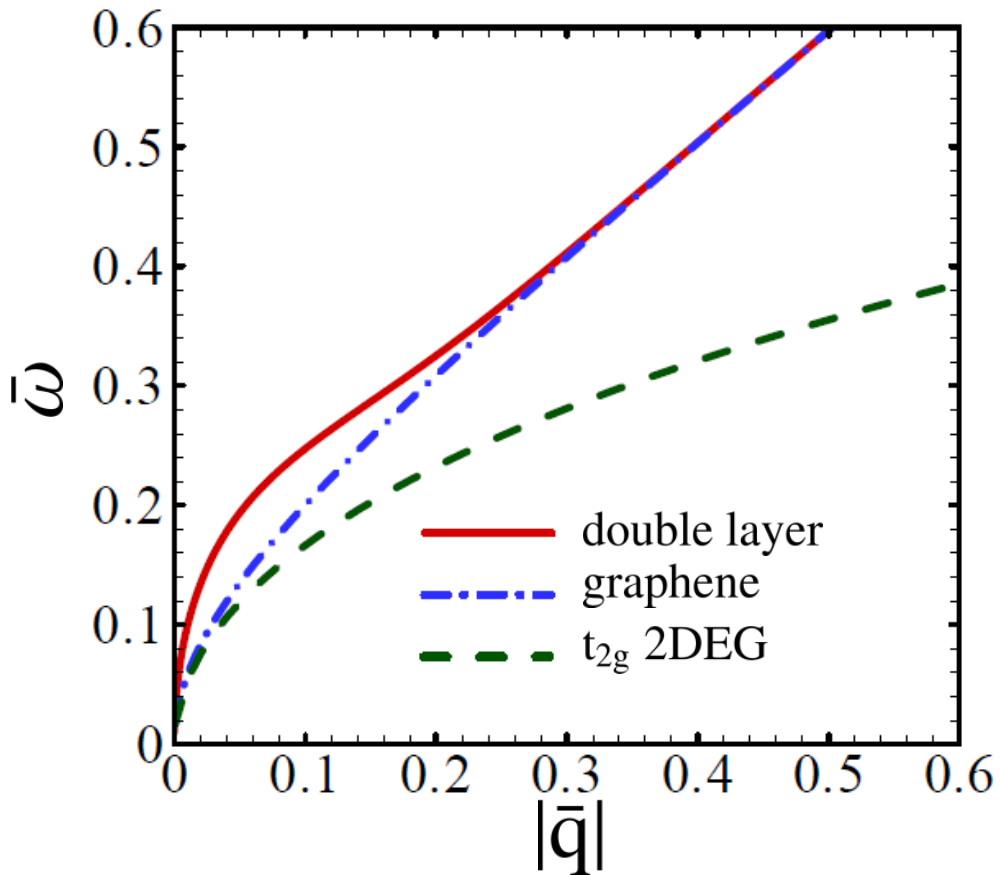}
 \caption{\label{fig12} (Color online) Optical modes of the decoupled double layer structure,  $\bar{\omega}(\bar{q})$ in comparison with that of graphene and $\rm{t_{2g}}$ 2DEG systems as a function of $|\bar{\mathbf{q}}|$ with the electron density $n_s=\rm{7\times 10^{13}~cm^{-2}}$ at the interface .}
\end{figure}

In Fig.~\ref{fig10} we illustrate all collective modes of the system for different values of electron density at the interface. The lower graphs in the figure show the acoustic modes and the upper graphs belong to optical modes of the system. The dashed line also shows the electron-hole continuum of graphene layer (which has the fastest carriers in the system). As seen from the figure, by lowering the electron density of the interface, the damping of acoustic plasmon starts in a smaller wave vector. The lower density of the interface results in weaker screening of the electrons of the bottom layer, so the Coulomb interaction between layers increases and in this case it is hard for the carriers of the layers to oscillate out-of-phase. In the inset, there are two other damped acoustic plasmons of the system for $n_s=\rm{7\times10^{13}~cm^{-2}}$ with green dashed lines specifying the electron-hole continuum of the three bands of the 2DEG at the interface.

In Fig.~\ref{fig11} we compare the collective modes of the decoupled system with the one with a simple one band 2DEG at the interface with $m^*=m_L$ and equal electron density as $\rm{t_{2g}}$ 2DEG case. The figure is plotted for two interface electron densities $n_s=\rm{7\times10^{13}~cm^{-2}}$ and $n_s=\rm{2\times10^{14}~cm^{-2}}$. We see that for the smaller electron density at the interface the optical plasmon dose not change significantly, but the acoustic mode starts damping in a smaller wave vector in the case of $\rm{t_{2g}}$ 2DEG. Increasing the electron density to $n_s=\rm{2\times10^{14}~cm^{-2}}$, not only we see the same behavior of the acoustic mode as the previous case, but also a considerable change in optical mode can be seen as well. As we increase the density of carriers at the interface, the $\rm{d_{xz}}$ and $\rm{d_{yz}}$ bands begin to be populated. In lower interface density regime the ratio of the carriers in these heavier bands is less than the isotropic and light $\rm{d_{xy}}$ band, therefore the behavior of the optical mode does not change significantly. But in higher interface densities the role of these bands becomes more and more important and their heavier effective mass causes damping in lower wave vectors. On the other hand, as the acoustic mode is a lower energy collective mode it is more sensitive to the characteristics of the system and even for lower densities at the interface, the less populated $\rm{d_{xz}}$ and $\rm{d_{yz}}$ bands can affect the acoustic mode considerably.

Finally, we compare the optical plasmon mode of the decoupled double layer system with that of graphene and ${\rm t}_{2g}$ 2DEG systems and the results are shown in Fig.~\ref{fig12}. The impact of correlations in the decoupled system is significantly large at the small and mid wave vector values. Although the optical mode of the ${\rm t}_{2g}$ 2DEG system grows slowly, the optical mode of graphene attains to the optical mode of the decoupled system at larger $q$ values. Accordingly, graphene and ${\rm t}_{2g}$ 2DEG are sensitive to their environment and their physical properties change by other materials surrounding them.

\section{Summary and Conclusions}

In the first part of this paper, we have investigated the electronic collective modes of the 2DEG residing at the interface of $\rm{LaAlO_3}$ and $\rm{SrTiO_3}$. Working in carrier density regime high enough to be able to forget the spin-orbit interaction, we have used a simple three-band Hamiltonian. Therefore, we expect to have three collective modes in this system, namely one optical mode and two acoustic modes.

We have demonstrated that while the low-lying acoustic mode is always damped, the emergence of the upper acoustic mode above the particle-hole excitations of the system, depends on the dielectric constant of the surrounding medium as well as the carrier density at the interface. We want to emphasize that considering the effective distance between circular and elliptical bands, is the key point in order to have an undamped acoustic mode at the long wavelength limit.
Note that lowering the electron density of the system, so that the electrons only occupy the states of the circular $\rm{d_{xy}}$ band, the system transforms to a simple 2DEG which has no more than the ordinary optical plasmon. Therefore the acoustic modes are expected to emerge in higher carrier density.

We have also discussed that because the fastest carriers of the system belong to the isotropic $\rm{d_{xy}}$ band, the anisotropy of the other two bands can not affect the collective modes of the system considerably and for any practical carrier density and band offset energy between $\rm{d_{xy}}$ and elliptical bands, the carriers of the former are the fastest carriers in the system.

In addition, we have derived analytical expressions for both optical and acoustic (damped and undamped) collective modes at the long wavelength limit which were in perfect agreement with the numerical findings.

The second part of this paper was devoted to the behavior of the collective modes of graphene-LAO/STO double layer separated by a distance $d$ in $\hat{z}$ direction which is the thickness of $\rm{LaAlO_3}$ layer in this case. In general, in a double layer system we have two collective modes, but the multiband nature of $\rm{t_{2g}}$ 2DEG increases the number of the plasmons, so that in this double layer structure there are  one optical mode and three acoustic modes two of which are completely damped as we have illustrated numerically.

Here again we have derived analytical expressions in the limit of small $qd$ for both optical and the top outermost acoustic plasmons. We have found an expression for the critical value of the interlayer distance above which the upper acoustic mode can emerge in the region where the imaginary part of the dielectric part of the dielectric function of the system vanishes and the mode is undamped at least  in the long wavelength limit.

We have compared our results with the case in which only the lowest band of the $\rm{t_{2g}}$ 2DEG is occupied. The faster damping of both optical and acoustic modes is concluded especially for higher electron density at the interface, where the elliptical bands contribution in the dielectric function and its effect on the plasmon modes of the system is increased.

The critical distance for the $\rm{t_{2g}}$ electron gas in comparison with the conventional 2DEG can have interesting consequences as well. If we have a double layer system with interlayer distance between the critical distance of the conventional 2DEG and that of $\rm{t_{2g}}$ electron gas, going from a one band to a multi band phase by increasing the electron density at the interface (neglecting the spin-orbit interaction), the acoustic plasmon is expected to disappear or at least weaken for the electron densities larger than a critical value(where $\rm{d_{xz}}$ and $\rm{d_{yz}}$ orbitals begin populating) and then again increasing the electron density results in a shaper acoustic mode .

We should note that there are still some other intrinsic damping mechanisms such as phonons and impurities which were not considered in our analysis. As we were working in a very low temperature regime, neglecting the lattice phonon scattering is a good approximation, but the influence of impurities on the plasmon modes especially in the case of not very clean interfaces can be of importance.

\section{Acknowledgement}

We thank A. Langari and F. Parhizgar for useful discussions. This work is partially supported by the Iran Science Elites Federation grant.

\appendix
\section{Long wavelength analytical expressions for the second acoustic plasmon of LAO/STO system}
\label{Sect:AppendixA}

To find analytical expressions for the group velocity and damping of the second acoustic mode of LAO/STO, we follow the same procedure described to get  Eqs.~\eqref{eq:delta} and  \eqref{eq:cs1} but as mentioned in the text, in this case the acoustic mode lies in the region $qv_{F2}z_s(\theta)<\omega<qv_{F2}z_m(\theta)$ with $z_s=min\lbrace z_1,z_2 \rbrace$  and $z_m=max\lbrace z_1,z_2 \rbrace$. After some lengthy but straightforward calculations, we arrive at
\begin{widetext}
\begin{equation}\label{eq:delta2}
\delta=-\frac{ABC[\sqrt{BC}(1+4k_2a)+\sqrt{AC}(m_D/m_L+2k_2a)-2k_2a\sqrt{B}c_s]}{(AB)^{3/2}z_2v_{F2}(m_D/m_L+2k_2a)c_s-2k_2aC^{3/2}(z_1v_{F2}A+v_{F1}B)c_s^2}
\end{equation}

and
\begin{equation}\label{eq:cs2}
 \begin{split}
&(2k_2a\delta B^{3/2})\Bigl[v_{F1}C-z_2v_{F2}A\Big]c_s^3\\
&-\Big[(BC)^{3/2}\delta v_{F1}(1+4k_2a)+AC^{3/2}(\delta z_1v_{F2}\frac{m_D}{m_L}\sqrt{A}+2k_2aB) \Big]c_s^2\\
&-(AB)^{3/2}C(\frac{m_D}{m_L}+2k_2a)c_s\\
&+(ABC)^{3/2}(1+2\frac{m_D}{m_L}+4k_2a)=0
 \end{split}
 \end{equation}
\end{widetext}
Substituting  Eq.~\eqref{eq:delta2} in  Eq.~\eqref{eq:cs2} and solving this equation for $c_s$, the group velocity of the second acoustic plasmon is found. Now putting $c_s$ back in expression \eqref{eq:delta2}, we will have the Landau damping of this mode as well. Note that the definitions of $A$, $B$, $C$ and $k_2$ are given after Eq.~12.


\begin{thebibliography}{7}

\bibitem{pines}
D. Pines and P. Nozier,
\textit{The Theory of Quantum Liquids }
 (W. A. Benjamin, New York, 1996).

 \bibitem{vignale}
G. F. Giuliani and G. Vignale,
\textit{Quantum Theory of the Electron Liquid }
 (Cambridge University Press, Cambridge, 2005).


\bibitem{pellegrini_review_2006}
 V. Pellegrini and A. Pinczuk,  Phys. Stat. Sol. (B) {\bf 243}, 3617 (2006); F. J. G. De Abajo
Rev. Mod. Phys. {\bf 82}, 209 (2010).

\bibitem{kainth_prb_1999}
	D. S. Kainth, D. Richards, H.P. Hughes, M.Y. Simmons, and D.A. Ritchie, \prb  {\bf 57}, R2065 (1998);
	D. S. Kainth, D. Richards, A. S. Bhatti, H.P. Hughes,
	M. Y. Simmons, E. H. Linfield, and D. A. Ritchie, {\it ibid.} {\bf 59}, 2095 (1999);	
	D. S Kainth, D. Richards, H. P. Hughes, M. Y. Simmons, and D. A. Ritchie,
	J. Phys.: Condens. Matter {\bf 12}, 439 (2000).

\bibitem{hirjibehedin_prb_2007}
	C. F. Hirjibehedin, A. Pinczuk, B. S. Dennis, L. N. Pfeiffer, and K.W. West, \prb {\bf 65}, 161309 (2002).

\bibitem{tovstonog_prb_2002}
	S. V. Tovstonog, L. V. Kulik, I.V. Kukushkin, A. V. Chaplik,
	J. H. Smet, K. V. Klitzing, D. Schuh, and G. Abstreiter, \prb {\bf 66}, 241308 (2002).	

\bibitem{liu_prb_2008}
	Y. Liu, R. F. Willis, K. V. Emtsev, and T. Seyller, \prb {\bf 78}, 201403(R) (2008).

\bibitem{eeinteractionsgraphene}
A. Bostwick, F. Speck, T. Seyller, K. Horn, M. Polini, R. Asgari, A.H. MacDonald, and E. Rotenberg, Science {\bf 328}, 999 (2010);
	A. Bostwick, T. Ohta, T. Seyller, K. Horn, and E. Rotenberg, Nature Phys. {\bf 3}, 36 (2007);
	Z.Q. Li, E.A. Henriksen, Z. Jiang, Z. Hao, M.C. Martin, P. Kim, H.L. Stormer, and D.N. Basov, {\it ibid.} {\bf 4}, 532 (2008);
	X. Du, I. Skachko, F. Duerr, A. Luican, and E.Y. Andrei, Nature (London) {\bf 462}, 192 (2009);
	K. I. Bolotin, F. Ghahari, M. D. Shulman, H. L. Stormer, and P. Kim, {\it ibid.} {\bf 462}, 196 (2009);
	V. W. Brar, S. Wickenburg, M. Panlasigui, C.-H. Park, T. O. Wehling,
	Y. Zhang, R. Decker, C. Girit, A. V. Balatsky, S. G. Louie, A. Zettl, and M. F. Crommie, \prl {\bf 104}, 036805 (2010);
	E. A. Henriksen, P. Cadden-Zimansky, Z. Jiang,
	Z. Q. Li, L.-C. Tung, M.E. Schwartz, M. Takita, Y.-J. Wang, P. Kim, and H. L. Stormer, {\it ibid.} {\bf 104}, 067404 (2010);
	A. Luican, G. Li, and E.Y. Andrei, \prb {\bf 83}, 041405(R) (2011);
	K.F. Mak, J. Shan, and T.F. Heinz, \prl {\bf 106}, 046401 (2011);
	F. Ghahari, Y. Zhao, P. Cadden-Zimansky, K. Bolotin, and P. Kim, {\it ibid.} {\bf 106}, 046801 (2011).

\bibitem{peralta_apl_2002}
	X. G. Peralta, S. J. Allen, M. C. Wanke, N. E. Harff, J.A. Simmons,
	M. P. Lilly, J. L. Reno, P. J. Burke, and J. P. Eisenstein, \apl {\bf 81}, 1627 (2002).

\bibitem{graphene}
F. H. L. Koppens, D. E. Chang, and F. J. Garcia de Abajo, Nano Lett. {\bf 11}, 3370 (2011);  Z. Fei,  A. S. Rodin, G. O. Andreev, W. Bao, A. S. McLeod,
M.  Wagner,   L.M.  Zhang,   Z.  Zhao,   M.  Thiemens,   G.
Dominguez,  M. M.  Fogler,  A. H.  Castro  Neto,  C.N.  Lau,
F. Keilmann, and D. N. Basov, Nature (London) {\bf 487}, 82 (2012); A. Woessner,  M. B. Lundeberg, Y.  Gao, A. Principi, P.
Alonso-Gonz\`{a}lez, M. Carrega, K. Watanabe, T. Taniguchi,
G. Vignale, M. Polini, J. Hone, R. Hillenbrand, and F.H.L.
Koppens, Natture Mater. {\bf 14}, 421 (2015).	
	

\bibitem{pogrebinskii_1977}
	M. B. Pogrebinsky, Sov. Phys. Semicond. {\bf 11}, 372 (1977).

\bibitem{price_physicaB_1983}
	P. J. Price, Physica B {\bf 117}, 750 (1983).

\bibitem{graphenereviews}
    A. K. Geim, I. V. Grigorieva, Nature (London), {\bf 499},  419 (2013);
	A. K. Geim, Science {\bf 324}, 1530 (2009);
	A. H. Castro Neto, F. Guinea, N. M. R. Peres, K.S. Novoselov, and A. K. Geim, \rmp {\bf 81}, 109 (2009);
	A. K. Geim and K. S. Novoselov, Nature Mater. {\bf 6}, 183 (2007).

\bibitem{ohtomo2}
 A. Ohtomo and H. Y. Hwang, Nature (London){\bf 427}, 423 (2004);

\bibitem{thiel}
 S. Thiel, G. Hammerl, A. Schmehl, C. W. Schneider, and
J. Mannhart, Science, {\bf 313}, 1942 (2006); H. Y. Hwang, Y. Iwasa, M. Kawasaki, B. Keimer, N. Nagaosa, and Y. Tokura, Nature Mater., {\bf 11}, 103 (2012).

\bibitem{Reyren}
N. Reyren, S. Thiel, A. D. Caviglia, L. F. Kourkoutis,
G. Hammerl, C. Richter, C. W. Schneider, T. Kopp, A. -
S. Ruetschi, D. Jaccard, M. Gabay, D. A. Muller, J. -
M. Triscone, and J. Mannhart, Science {\bf 317}, 1196 (2007);
  P. D. Eerkes, W. G. Van Der Wiel, and H. Hilgenkamp,
Appl. Phys. Lett., {\bf 103}, 201603 (2013).

\bibitem{Brinkman}
A. Brinkman, M. Huijben, M. van Zalk, J. Huijben,
U. Zeitler, J. C. Maan, W. G. van der Wiel, G. Rijnders,
D. H. A. Blank, and H. Hilgenkamp, Nature Mater. {\bf 6}, 493
(2007);
 D. Stornaiuolo, C. Cantoni, G. M. De Luca, R. Di Capua,
E. Di. Gennaro, G. Ghiringhelli, B. Jouault, D. Marr`e,
D. Massarotti, F. Miletto Granozio, I. Pallecchi, C. Piamonteze,
S. Rusponi, F. Tafuri, and M. Salluzzo, Nature Mater. {\bf 15}, 278 (2016).



\bibitem{Huang}
M. Huang, G. Jnawali, J. -F. Hsu, S. Dhingra, H. Lee,
S. Ryu, F. Bi, F. Ghahari, J. Ravichandran, L. Chen,
P. Kim, C. -B. Eom, B. D Urso, P. Irvin, and J. Levy,
APL Materials {\bf 3}, 062502 (2015).

\bibitem{Marco}
I. Aliaj, I. Torre, V. Miseikis, E. di Gennaro, A. Sambri, A. Gamucci, C. Coletti, F. Beltram, F. M. Granozio, M. Polini, V. Pellegrini, and S. Roddaro,  APL Mater. {\bf 4}, 066101 (2016).

 \bibitem{ruotsa}
K. O. Ruotsalainen \textit{et al}, J. Phys.: Condens. Matter {\bf27}, 335501 (2015).

\bibitem{Park}
Se Young Park and A. Millist, Phys. Rev. B {\bf 87}, 205145 (2013).

\bibitem{copie}
O. Copie, V. Garcia, C. B\"{o}defeld, C. Carr\'{e}t\'{e}ro, M. Bibes, G. Herranz, E. Jacquet, J.-L. Maurice, B. Vinter, S. Fusil, K. Bouzehouane, H. Jaffr\'{e}s, and A. Barth\`{e}l\`{e}my, Phys. Rev. Lett. {\bf 102}, 216804 (2009).

 \bibitem{gariglio}
S. Gariglio, A. Fete and J. M. Triscone, J. Phys.: Condens. Matter {\bf27}, 283201 (2015).

\bibitem{cancellieri}
C.  Cancellieri,  A.  S.  Mishchenko,  U.  Aschauer,  A.  Filippetti, C. Faber, O. S. Bariˇsi ́c, V. A. Rogalev, T. Schmitt,N. Nagaosa, and V. N. Strocov, Nat. Commun.7, 10386 (2016)

\bibitem{wang}
Z. Wang \textit{et al}, Nat. Mater.15,835 (2016)

\bibitem{kalabukhov}
A. Kalabukhovt, R. Gunnarsson, J.Borjesson, E. Olsson, T. Claeson, and D. Winkler, Phys. Rev. B {\bf 75},121404(R) (2007).

  \bibitem{herranz}
G. Herranz \textit{et al}, Phys. Rev. Lett{\bf 98}, 216803 (2007).

  \bibitem{sulpizio}
J. A. Sulpizio, S. Ilani, P. Irvin, and J. Levy, Ann. Rev. Mater. Res. 44, 117 (2014).

\bibitem{pentcheva}
R. Pentcheva and W. E. Pickett, Phys. Rev. B {\bf 74}, 035112 (2006).

\bibitem{zoran}
Z. S. Popovic, S. Satpathy and R.M. Martin, Phys. Rev. Lett. {\bf101}, 256801 (2008).


 \bibitem{khalsa}
G. Khalsa and A. H. MacDonald, Phys. Rev. B {\bf 86}, 125121 (2012).

  \bibitem{khalsa2}
G. Khalsa, B. Lee and A. H. MacDonald, Phys. Rev. B {\bf 88}, 041302 (2013).

  \bibitem{zhong}
Z. Zhong, A. T\'{o}th, and K. Held and F. Zhang, Phys. Rev. B {\bf 87}, 161105(R) (2013).

\bibitem{nayak}
Y. Kim, R. M. Lutchyn and C. Nayak, Phys. Rev. B {\bf 87}, 245121 (2013).

\bibitem{tolsma}
J. R. Tolsma, A. Principi, R. Asgari, M. Polini and A. H. MacDonald, Phys. Rev. B {\bf 93}, 045120 (2016).

\bibitem{tolsma2}
J. R. Tolsma, M. Polini and A. H. MacDonald, arXiv: 1608. 03625.

\bibitem{Faridi}
F. Faridi, R. Asgari, and A. Langari, Phys. Rev. B {\bf 93}, 235306 (2016).

\bibitem{zhou1}
J. Zhou, W.-Y. Shan, and D. Xiao, Phys. Rev. B {\bf 91}, 241302 (2015).

\bibitem{herring}
C. Herring and E. Vogt, Phys. Rev.{\bf101}, 944 (1956).

 \bibitem{dassarma2}
S. Das Sarma and J.J. Quinn, Phys. Rev. B {\bf 25}, 7603 (1982).


 \bibitem{takada}
Y. Takada, J. Phys.Soc. Jpn.{\bf 43}, 1627 (1977).

\bibitem{santoro}
G.E. Santoro and G.F. Giuliani, Phys. Rev. B {\bf 37}, 937 (1988).

 \bibitem{profumo}
R. E. V. Profumo, M. Polini, R. Asgari, R. Fazio and A. H. MacDonald, Phys. Rev. B {\bf 82}, 085443 (2010).

 \bibitem{e.h.hwang}
E. H. Hwang and S. Das Sarma, Phys. Rev. B {\bf 75}, 205418 (2007); T. Stauber
J. Phys.: Condens. Matter {\bf 26} 123201 (2014).


 \bibitem{dassarma1}
S. Das Sarma and A. Madhukar, Phys. Rev. B {\bf 23}, 805 (1981).

 \bibitem{rosario}
R. E.V. Profumo, R. Asgari, M. Polini and A. H. MacDonald, Phys. Rev. B {\bf 85}, 085443 (2012).

 \bibitem{principi}
A. Principi, M. Carrega, R. Asgari, V. Pellegrini and M. Polini, Phys. Rev. B {\bf 86}, 085421 (2012).

 \bibitem{stauber}
T. Stauber and G. Gomez-Santos, New J. Phys.{\bf 14}, 105018 (2012).

\end{thebibliography}
\end{document}